\documentclass[letterpaper,titlepage,11pt]{article}
\usepackage{hyperref}

\usepackage{amssymb,amsmath,amsfonts}
\usepackage{epsfig}

\def\clock{{\count0=\time
           \divide\count0 60
           \ifnum\count0<10 0\fi\the\count0
           \multiply\count0 -60 \advance\count0 \time
           :\ifnum\count0<10 0\fi \the\count0
         }}
\newcommand{\timestamp}{{\small\vbox{\hbox{\tt\jobname.tex}
\hbox{\the\day/\the\month/\the\year, \clock}}}}

\newcommand{\be}{\begin{equation}} \newcommand{\ee}{\end{equation}}
\newcommand{\bea}{\begin{eqnarray}} \newcommand{\eea}{\end{eqnarray}}

\newcommand{\CA}{\mathcal{A}}
\newcommand{\CC}{\mathcal{C}}

\newcommand{\CH}{\mathcal{H}}

\newcommand{\CO}{\mathcal{O}}

\newcommand{\CN}{\mathcal{N}}

\newcommand{\CV}{\mathcal{V}}

\newcommand{\id}{\hbox{1\kern-.27em l}}
\newcommand{\sid}{\hbox{\scriptsize1\kern-.27em l}}

\newcommand{\we}{\kern-.1em\wedge\kern-.1em}
\newcommand{\scal}{\kern-.13em\cdot\kern-.13em}

\newcommand{\II}{I\kern-.09em I}

\newcommand{\ads}{{\rm AdS}}

\newcommand{\spa}{\ , \ \ }

\newcommand{\tr}{\mathop{{\rm Tr}}}

\newcommand{\gym}{g_{\mathrm{YM}}}

\newcommand{\R}{\mathbb{R}}

\newcommand{\beastar}{\begin{eqnarray*}}
\newcommand{\eeastar}{\end{eqnarray*}}

\newcommand{\ds}{\displaystyle}

\numberwithin{equation}{section}

\begin{document}

\begin{titlepage}

\rightline{\vbox{\small\hbox{\tt hep-th/0605234} }}
\rightline{\vbox{\small\hbox{\tt Nordita-2006-14} }} \vskip 2.7cm

\centerline{\LARGE \bf Quantum Mechanical Sectors in} \vskip 2mm
\centerline{\LARGE \bf Thermal $\CN=4$ Super Yang-Mills on $\R
\times S^3$} \vskip 1.6cm

\centerline{\large {\bf  Troels Harmark} and {\bf Marta Orselli} }
\vskip 1.1cm
\begin{center}
\sl The Niels Bohr Institute and Nordita\\
\sl Blegdamsvej 17, 2100 Copenhagen \O , Denmark
\end{center}

\vskip 0.7cm

\centerline{\small\tt harmark@nbi.dk, orselli@nbi.dk}

\vskip 1.7cm

\centerline{\bf Abstract} \vskip 0.2cm \noindent We study the
thermodynamics of $U(N)$ $\CN=4$ Super Yang-Mills (SYM) on $\R
\times S^3$ with non-zero chemical potentials for the $SU(4)$
R-symmetry. We find that when we are near a point with zero
temperature and critical chemical potential, $\CN=4$ SYM on $\R
\times S^3$ reduces to a quantum mechanical theory. We identify
three such critical regions giving rise to three different quantum
mechanical theories. Two of them have a Hilbert space given by the
$SU(2)$ and $SU(2|3)$ sectors of $\CN=4$ SYM of recent interest in
the study of integrability, while the third one is the half-BPS
sector  dual to bubbling AdS geometries. In the planar limit the
three quantum mechanical theories can be seen as spin chains. In
particular, we identify a near-critical region in which $\CN=4$
SYM on $\R \times S^3$ essentially reduces to the ferromagnetic
$XXX_{1/2}$ Heisenberg spin chain. We find furthermore a limit in
which this relation becomes exact.


\end{titlepage}


\pagestyle{plain} \setcounter{page}{1}

\tableofcontents

\section{Introduction}

The thermodynamics of large $N$ $U(N)$ $\CN=4$ Super Yang-Mills
(SYM) on $\R \times S^3$ has proven to be interesting for several
reasons. It has a confinement/deconfinement phase transition like
in QCD that can be studied even at weak coupling
\cite{Witten:1998zw}. This phase transition is conjectured to
correspond to the Hagedorn phase transition for the dual type IIB
string theory on $\ads_5 \times S^5$, which is in accordance with
the fact that the large $N$ $\CN=4$ SYM theory has a Hagedorn
spectrum \cite{Sundborg:1999ue,Polyakov:2001af,Aharony:2003sx}.
This is very interesting since it means that we can study what
happens beyond the Hagedorn transition on the weakly coupled gauge
theory side. For large coupling the same phase transition
corresponds to the Hawking-Page phase transition for black holes
in Anti-De Sitter space, which is a phase transition in
semi-classical gravity \cite{Hawking:1982dh,Witten:1998zw}. Thus,
by studying the thermodynamics of $\CN=4$ SYM we can hope to learn
about such important subjects as what is beyond the Hagedorn
transition, confinement in QCD and phase transitions in gravity.

In this paper we find that thermal $U(N)$ $\CN=4$ SYM has quantum
mechanical sectors, by which we mean that near certain critical
points most of the degrees of freedom of $\CN=4$ SYM can be
integrated out and only a small subset, that we can regard as
quantum mechanical, remains. These critical points arise in the
study of the thermodynamics of $U(N)$ $\CN=4$ SYM on $\R \times
S^3$ with non-zero chemical potentials corresponding to the three
R-charges for the $SU(4)$ R-symmetry of $\CN=4$ SYM. Our main
result is that when we are near a point with zero temperature and
critical chemical potentials, $\CN=4$ SYM reduces to one out of
three simple quantum mechanical theories. Furthermore, for large
$N$ these three quantum mechanical theories are mapped in a
precise way to spin chain theories.

Denoting the three chemical potentials of $\CN=4$ SYM as
$\Omega_1$, $\Omega_2$, $\Omega_3$ and setting
$\Omega_1=\Omega_2=\Omega$, $\Omega_3=0$, we can write one of the
near-critical regions that we study as
\begin{equation}
\label{su2lim} T \ll 1 \spa 1-\Omega \ll 1 \spa \lambda \ll 1
\end{equation}
where $T$ is the temperature and $\lambda$ is the 't Hooft
coupling of $\CN=4$ SYM. In this region we are close to the
critical point $(T,\Omega_1,\Omega_2,\Omega_3)=(0,1,1,0)$. We show
in this paper that in the region \eqref{su2lim} $\CN=4$ SYM on $\R
\times S^3$ reduces to a quantum mechanical theory with the
Hilbert space consisting of all multi-trace operators made out of
the letters $Z$ and $X$, where $Z$ and $X$ are complex scalars of
$\CN=4$ SYM with R-symmetry weights $(1,0,0)$ and $(0,1,0)$. This
is precisely the so-called $SU(2)$ sector that has been discussed
in recent developments on the integrability of $\CN=4$ SYM
\cite{Minahan:2002ve,Beisert:2003tq,Beisert:2003jj,Beisert:2003yb,Beisert:2004hm}.

We find that it is natural to reformulate $\CN=4$ SYM in the
region \eqref{su2lim} in terms of the rescaled temperature
$\tilde{T} \equiv T/(1-\Omega)$. Writing the dilatation operator
of $\CN=4$ SYM as $D = D_0 + \lambda D_2 + \CO (\lambda^2)$ where
$D_0$ is the zeroth order dilatation operator and $\lambda$ is the
't~Hooft coupling, we can write the leading terms of the
Hamiltonian of our quantum mechanical theory as
\begin{equation}
\label{theH} H = D_0 + \tilde{\lambda} D_2
\end{equation}
where $\tilde{\lambda} \equiv \lambda /(1-\Omega)$ is a rescaled
coupling. This resembles the leading terms of the dilatation
operator of the $SU(2)$ sector except for the rescaled coupling
$\tilde{\lambda}$. The first correction to \eqref{theH} is of
order $\tilde{\lambda} \lambda$. Our result is thus that in the
near-critical region \eqref{su2lim} $\CN=4$ SYM on $\R \times S^3$
reduces to a quantum mechanical theory with temperature
$\tilde{T}$, Hamiltonian \eqref{theH} (for the leading terms) and
with the Hilbert space corresponding to the $SU(2)$ sector of
$\CN=4$ SYM.

For large $N$ we can focus on single-trace operators of a certain
length $L$. Such operators can be thought of as periodic spin
chains of length $L$. The Hamiltonian \eqref{theH} is then $L +
\tilde{\lambda} D_2$ and $D_2$ is known to correspond to the
ferromagnetic $XXX_{1/2}$ Heisenberg spin chain Hamiltonian. Thus,
for $N=\infty$ our result is that thermal $\CN=4$ SYM on $\R
\times S^3$ reduces to the ferromagnetic $XXX_{1/2}$ Heisenberg
spin chain, in the sense that we have a precise relation between
the partition functions of the two theories.

A further result of this paper is that if we take the limit
\begin{equation}
\label{exactlim} T \rightarrow 0 \spa \Omega \rightarrow 1 \spa
\lambda \rightarrow 0 \spa \tilde{T} = \frac{T}{1-\Omega} \
\mbox{fixed} \spa \tilde{\lambda}= \frac{\lambda}{1-\Omega}\
\mbox{fixed}
\end{equation}
the Hamiltonian \eqref{theH} becomes exact with the Hilbert space
being the $SU(2)$ sector. Hence, for $N=\infty$ and in the limit
\eqref{exactlim} we have that the relation between the partition
function of $\CN=4$ SYM on $\R \times S^3$ and that of the
ferromagnetic $XXX_{1/2}$ Heisenberg spin chain is exact, i.e. we
find that
\begin{equation}
\log Z_{\CN=4} (\tilde{T}) = \sum_{n=1}^\infty \sum_{L=1}^\infty
\frac{1}{n} e^{-nL/\tilde{T}} Z^{(XXX)}_L (\tilde{T}/n)
\end{equation}
where $Z_{\CN=4}$ is the partition function for $\CN=4$ SYM on $\R
\times S^3$ and $Z^{(XXX)}_L$ is the partition function for the
ferromagnetic $XXX_{1/2}$ Heisenberg spin chain of length $L$ with
Hamiltonian $\tilde{\lambda} D_2$.

We consider furthermore two other near-critical regions. Near
$(T,\Omega_1,\Omega_2,\Omega_3)=(0,1,1,1)$ we find that $\CN=4$
SYM on $\R \times S^3$ reduces to a quantum mechanical theory in
the so-called $SU(2|3)$ sector of $\CN=4$ SYM which also recently
has been considered in the study of integrability
\cite{Beisert:2003jj, Beisert:2003ys}. This sector consists of
three complex scalars and two complex fermions. We find similar
results in this sector as for the $SU(2)$ sector.

Near $(T,\Omega_1,\Omega_2,\Omega_3)=(0,1,0,0)$ we find instead
that $\CN=4$ SYM on $\R \times S^3$ reduces to the half-BPS sector
consisting of multi-trace operators made of a single complex
scalar $Z$. This sector is precisely the half-BPS sector dual to
the bubbling $\ads$ geometries of \cite{Lin:2004nb}. As part of
this, it also contains the states dual to the vacuum of the
maximally supersymmetric pp-wave background
\cite{Blau:2001ne,Berenstein:2002jq}, to $AdS_5 \times S^5$
\cite{Aharony:1999ti}, and to giant gravitons in $AdS_5 \times
S^5$ \cite{McGreevy:2000cw}. The reduction of $\CN=4$ SYM to the
half-BPS sector was previously considered in \cite{Yamada:2005um}.

Finally, we consider the one-loop partition function for planar
$\CN=4$ SYM on $\R \times S^3$ with non-zero chemical potentials
and we find the corrected Hagedorn temperature, generalizing
\cite{Spradlin:2004pp}. We find furthermore the explicit form of
the corrected partition functions and Hagedorn temperature for the
$SU(2)$ and $SU(2|3)$ sectors. As a consistency check, we verify
that one gets the same result by taking the limit of the full
partition function as what one gets from the reduced partition
functions.

This paper is structured as follows. In Section
\ref{sec:freethermal} we consider free $\CN=4$ SYM on $\R\times
S^3$. We compute the partition function with non-zero chemical
potentials in Section \ref{sec:partfct} and we find the Hagedorn
temperature in Section \ref{sec:hagtemp}. In Section
\ref{sec:nearcrit} we identify the three near-critical regions and
we show the reductions to the half-BPS sector, the $SU(2)$ sector
and the $SU(2|3)$ sector. We consider furthermore these reductions
in the oscillator basis of $\CN=4$ SYM in Appendix \ref{app:osc}.
Finally in Section \ref{sec:aboveTH} we consider the
thermodynamics above the Hagedorn temperature.

In Section \ref{sec:QMsec} we consider the three near-critical
regions for interacting $\CN=4$ SYM on $\R\times S^3$ and find
that we still have the reductions to the half-BPS sector, the
$SU(2)$ sector and the $SU(2|3)$ sector, but now with a
non-trivial Hamiltonian. For $N=\infty$ we relate this Hamiltonian
to spin chain Hamiltonians, in particular we find that the $SU(2)$
sector has a Hamiltonian with the leading part given by the
ferromagnetic $XXX_{1/2}$ Heisenberg spin chain. We briefly review
the $XXX_{1/2}$ Heisenberg spin chain in Appendix \ref{app:XXX}.

In Section \ref{sec:lowtemp} we consider the low temperature limit
for the near-critical region in which $\CN=4$ SYM reduces to the
$SU(2)$ sector. In this case we find for large $N$ that the
ferromagnetic $XXX_{1/2}$ Heisenberg spin chain governs the
dynamic and from this we can find which states we are driven
towards as we take the temperature to zero.

In Section \ref{sec:declim} we write down the decoupling limit
mentioned above, from which it follows for the $SU(2)$ sector that
we have an exact relation between $\CN=4$ SYM and the $XXX_{1/2}$
Heisenberg spin chain for $N=\infty$.

In Section \ref{sec:oneloop} we consider the one-loop correction to
the thermal partition function of large $N$ $U(N)$ $\CN=4$ SYM on
$\R\times S^3$. We show how to compute the partition function with
non-zero chemical potentials, following \cite{Spradlin:2004pp}. We
have put part of this computation in Appendix \ref{app:oneloop}. We
find the one-loop corrected Hagedorn temperature both for small
chemical potential and near the critical points. Near the critical
points we also find the partition function explicitly, and we find
that the one-loop partition function of $\CN=4$ SYM on $\R \times
S^3$ indeed correctly reduces to the one of the reduced theories.

In Section \ref{sec:concl} we present our conclusions and discuss
future directions.

\noindent {\sl Note on related work:} We note that during the work
on this paper the article \cite{Yamada:2006rx} appeared with
results that overlap with Sections \ref{sec:partfct} and
\ref{sec:hagtemp}.

\section{Free thermal $\CN=4$ SYM on $\R \times S^3$}
\label{sec:freethermal}

We consider in this section the thermal partition function of
$\CN=4$ SYM on $\R \times S^3$ with chemical potentials at zero
coupling.

\subsection{Calculation of the partition function}
\label{sec:partfct}

In this section we consider the generalization of the computation
of the partition function for $U(N)$ $\CN=4$ SYM on $\R \times
S^3$ at zero coupling $\gym^2 = 0$ in
\cite{Sundborg:1999ue,Polyakov:2001af,Aharony:2003sx} to include
the three chemical potentials associated with the $SU(4)$
R-symmetry of $\CN=4$ SYM.

The partition function of $U(N)$ $\CN=4$ SYM on $\R \times S^3$ is
given by the trace of $e^{-\beta H}$ over all of the physical
states, where $\beta=1/T$ is the inverse temperature and $H$ is
the Hamiltonian. From the state/operator correspondence we have
that any state of $U(N)$ $\CN=4$ SYM on $\R \times S^3$ can be
mapped to a gauge invariant operator of $U(N)$ $\CN=4$ SYM on
$\R^4$. The Hamiltonian is then mapped to the dilatation operator
$D$ (here and in the following we set the radius of $S^3$ to one).
The Gauss constraint for a $U(N)$ gauge theory on $\R \times S^3$
means that we can only have states which are singlets of $U(N)$.
For operators, this means that the set of operators consists of
multi-trace operators made by combining single-trace operators,
where each single-trace operator is made from combining individual
letters, a letter being any operator one can make using a single
field of $\CN=4$ SYM and the covariant derivative
\cite{Sundborg:1999ue,Polyakov:2001af,Aharony:2003sx}.

To include the chemical potential associated with the $SU(4)$
R-symmetry of $\CN=4$ SYM we need to introduce the R-charges. Let
$R_1$, $R_2$ and $R_3$ denote the Cartan generators of $SU(4)$
(corresponding to the standard Cartan generators of $SO(6)$). Then
$R_1$, $R_2$ and $R_3$ are the three R-charges of $\CN=4$ SYM and
corresponding to these we have three chemical potentials
$\Omega_1$, $\Omega_2$ and $\Omega_3$. When computing a partition
function in the grand canonical ensemble one should compute the
trace of $e^{-\beta H + \beta \Omega_1 R_1+ \beta \Omega_2 R_2+
\beta \Omega_2 R_2}$ over all the physical states.

For the free $\CN=4$ SYM theory we should use the zeroth order
dilatation operator $D_0$ as the Hamiltonian. We can then
schematically write the full partition function in the grand
canonical ensemble as
\begin{equation}
Z(x,y_1,y_2,y_3) = {\tr}_M \left( x^{D_0} y_1^{R_1} y_2^{R_2}
y_3^{R_3} \right)
\end{equation}
Here we write $M$ for the set of multi-trace operators (or rather
the corresponding states) and we introduce the useful book keeping
devices
\begin{equation}
x \equiv e^{-\beta} \spa y_i \equiv e^{\beta \Omega_i} \ , \
i=1,2,3
\end{equation}
We note the important point that for finite $N$ not all
multi-trace operators are linearly independent. Certain
single-trace operators can for example by written in terms of
multi-trace operators. We therefore assume $M$ to be defined such
that all of the multi-trace operators in $M$ are linearly
independent, since otherwise we would count too many states
\cite{Sundborg:1999ue,Aharony:2003sx}.

To compute the partition function one should first find the
partition function for a single letter. To do this, we need to
understand the possible letters one can have and what their
conformal dimensions and R-charges are. The field content of
$\CN=4$ SYM consists of 6 real scalars $\phi_a$, $a=1,...,6$, a
gauge boson $A_\mu$ and the complex fermionic fields
$\psi^\alpha_A$, $\bar{\psi}_{\dot{\alpha}}^A$,
$\alpha,\dot{\alpha}=1,2$, $A=1,2,3,4$, corresponding to 16 real
fermionic components. The scalars all have conformal dimension 1,
the gauge boson also have dimension one while the fermions have
dimension $3/2$. With respect to the $SU(4)$ R-symmetry we have
that the 6 scalars correspond to a $[0,1,0]$ representation, the
gauge boson is a singlet under $SU(4)$ R-symmetry, while the
fermions correspond to a $[1,0,0]$ and a $[0,0,1]$ representation
of $SU(4)$. With respect to $(R_1,R_2,R_3)$ we then have that for
instance the $[0,1,0]$ representation corresponding to the 6
scalars have weights $(\pm 1,0,0)$, $(0,\pm 1,0)$ and $(0,0,\pm
1)$. For use in following sections of this paper we define here
the three complex scalars $Z=\phi_1+i\phi_2$, $X=\phi_3+i\phi_4$
and $W=\phi_5+i\phi_6$, corresponding to the weights $(1,0,0)$,
$(0,1,0)$ and $(0,0,1)$, respectively.

The set of letters of $\CN=4$ SYM, here denoted by $\CA$, is the
set of all the different operators on $\R^4$ that one can form by
applying the covariant derivative an arbitrary number of times on
either one of the scalars $\phi_a$, on the gauge field strength
$F_{\mu\nu}$ or on one of the fermions $\psi^\alpha_A$,
$\bar{\psi}_{\dot{\alpha}}^A$. These operators should be
independent of each other in the sense that two operators which
are related by the EOMs count as the same operator. It is well
known \cite{Sundborg:1999ue,Polyakov:2001af,Aharony:2003sx} that a
scalar on $\R \times S^3$ has letter partition function
$(x+x^2)/(1-x)^3$, a fermion $2x^{3/2}/(1-x)^3$ and a gauge boson
$(6x^2-2x^3)/(1-x)^3$. Using this, we get the following letter
partition function for $\CN=4$ SYM on $\R \times S^3$
\begin{equation}
\label{lettpart}
\begin{array}{l}
z(x,y_1,y_2,y_3) = {\tr}_\CA \left( x^{D_0} y_1^{R_1} y_2^{R_2}
y_3^{R_3} \right) \\[2mm] \ds = \frac{6x^2-2x^3}{(1-x)^3} +
\frac{x+x^2}{(1-x)^3} \sum_{i=1}^3 \left( y_i+y_i^{-1} \right)  +
\frac{2x^{3/2}}{(1-x)^3} \prod_{i=1}^3
\left(y_i^{\frac{1}{2}}+y_i^{-\frac{1}{2}}\right)
\end{array}
\end{equation}

If we consider the large $N$ case, we can for small enough
energies $E \ll N^2$ ignore the non-trivial relations between
multi-trace operators, e.g. the set of single-trace operators is
well-defined in this case. This enables us to make a purely
combinatorical computation of the partition function. One begins
by computing the single-trace partition function. The single trace
operators are $\tr (A_1 A_2 \cdots A_L )$ with $A_i \in \CA$. Note
that here and in the following we take the $U(N)$ trace to be in
the adjoint representation of $U(N)$. One can then use standard
combinatorical techniques to find the single-trace partition
function as \cite{Sundborg:1999ue,Polyakov:2001af,Aharony:2003sx}
\begin{equation}
\label{ZST} Z_{\rm ST} (x,y_1,y_2,y_3) = - \sum_{k=1}^\infty
\frac{\varphi(k)}{k} \log \left[
1-z(\omega^{k+1}x^k,y_1^k,y_2^k,y_3^k) \right]
\end{equation}
where we introduced the useful quantity $\omega=e^{2\pi i}$ which
is $-1$ if uplifted to a half-integer power, following
\cite{Spradlin:2004pp}. In this way we ensure that the fermionic
part of the partition function has the correct sign corresponding
to fermionic statistics. In \eqref{ZST} $\varphi(k)$ is the Euler
totient function which appears here due to the combinatorical
complication that the single-trace operators have a cyclic
symmetry.

The complete partition function $Z(x,y_1,y_2,y_3)$ for $U(N)$
$\CN=4$ SYM on $\R \times S^3$ with $N=\infty$, which traces over
all the multi-trace operators build from the single-trace
operators, can then be found as
\begin{equation}
\label{multinfty} \begin{array}{rcl} \log Z(x,y_1,y_2,y_3) &=& \ds
\sum_{n=1}^\infty \frac{1}{n} Z_{\rm ST} \left(
\omega^{n+1}x^{n},y_1^n,y_2^n,y_3^n \right) \\[2mm] &=& \ds -
\sum_{k=1}^\infty \log \left[ 1 - z(\omega^{k+1}
x^k,y_1^k,y_2^k,y_3^k) \right]
\end{array}
\end{equation}

By a more careful analysis one can find the partition function for
finite $N$, in which case there are non-trivial relations between
the multi-trace operators. The partition function for $U(N)$
$\CN=4$ SYM on $\R \times S^3$ with chemical potentials is
\cite{Aharony:2003sx,Basu:2005pj}
\begin{equation}
\label{fullZN} Z (x,y_1,y_2,y_3) = \int [dU] \exp \left[
\sum_{k=1}^\infty \frac{1}{k} z(\omega^{k+1}
x^k,y_1^k,y_2^k,y_3^k) \tr ( U^k ) \tr ( (U^\dagger)^k ) \right]
\end{equation}
Here $\int [dU] $ is the integral over the group $U(N)$ normalized
such that $\int [dU] = 1$. As mentioned above, we take the trace
over $U(N)$ to be in the adjoint representation.

\subsection{Hagedorn temperature for non-zero chemical potentials}
\label{sec:hagtemp}

If we consider the $N = \infty$ partition function
Eq.~\eqref{multinfty} for $U(N)$ $\CN=4$ SYM on $\R \times S^3$
it is clear that there is a singularity when
\begin{equation}
\label{hageq} z(x,y_1,y_2,y_3) = 1
\end{equation}
This is the Hagedorn singularity of the partition function
\eqref{multinfty}
\cite{Sundborg:1999ue,Polyakov:2001af,Aharony:2003sx} here
generalized to include non-zero chemical potentials. It is easy to
see that \eqref{hageq} with \eqref{lettpart} for given chemical
potentials $\Omega_i$ defines a critical temperature
$T_H(\Omega_1,\Omega_2,\Omega_3)$. One can check from the
partition function \eqref{multinfty} that there are no
singularities for $T < T_H(\Omega_1,\Omega_2,\Omega_3)$.

For temperatures just below the Hagedorn temperature, write
\begin{equation}
z(x,y_1,y_2,y_3) = 1 - \frac{T_H-T}{T_H C} + \CO ( (T_H-T)^2 )
\end{equation}
for $0 \leq T_H(\Omega_1,\Omega_2,\Omega_3)-T \ll
T_H(\Omega_1,\Omega_2,\Omega_3)$ with
$C=C(\Omega_1,\Omega_2,\Omega_3)$. Then the partition function for
temperatures just below the Hagedorn temperature has the behavior
\begin{equation}
\label{Zsingu} Z(T,\Omega_1,\Omega_2,\Omega_3) \simeq \frac{T_H
C}{T_H-T}
\end{equation}
From this one can find that the density of states for single-trace
operators is $E^{-1} e^{E/T_H}$ \cite{Aharony:2003sx}. Therefore,
when $N=\infty$ we have a Hagedorn density of states for large
energies.

\begin{figure}[ht]
\centerline{\epsfig{file=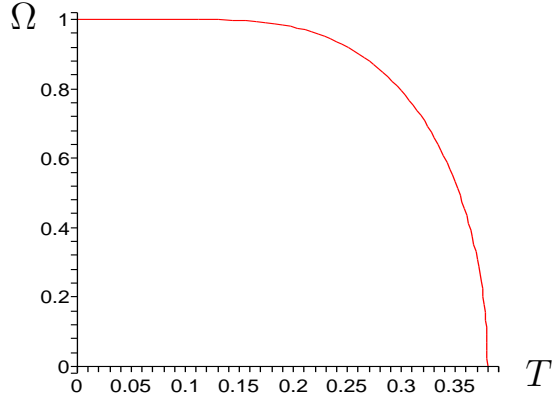,width=7cm,height=6cm} }
\caption{The Hagedorn temperature $T_H$ as function of $\Omega$ in
the case
$(\Omega_1,\Omega_2,\Omega_3)=(\Omega,0,0)$.\label{figTO1}}
\begin{picture}(0,0)(0,0)
\put(115,190){\Large $\Omega$} \put(310,55){\Large $T$}
\end{picture}
\end{figure}

\begin{figure}[ht]
\centerline{\epsfig{file=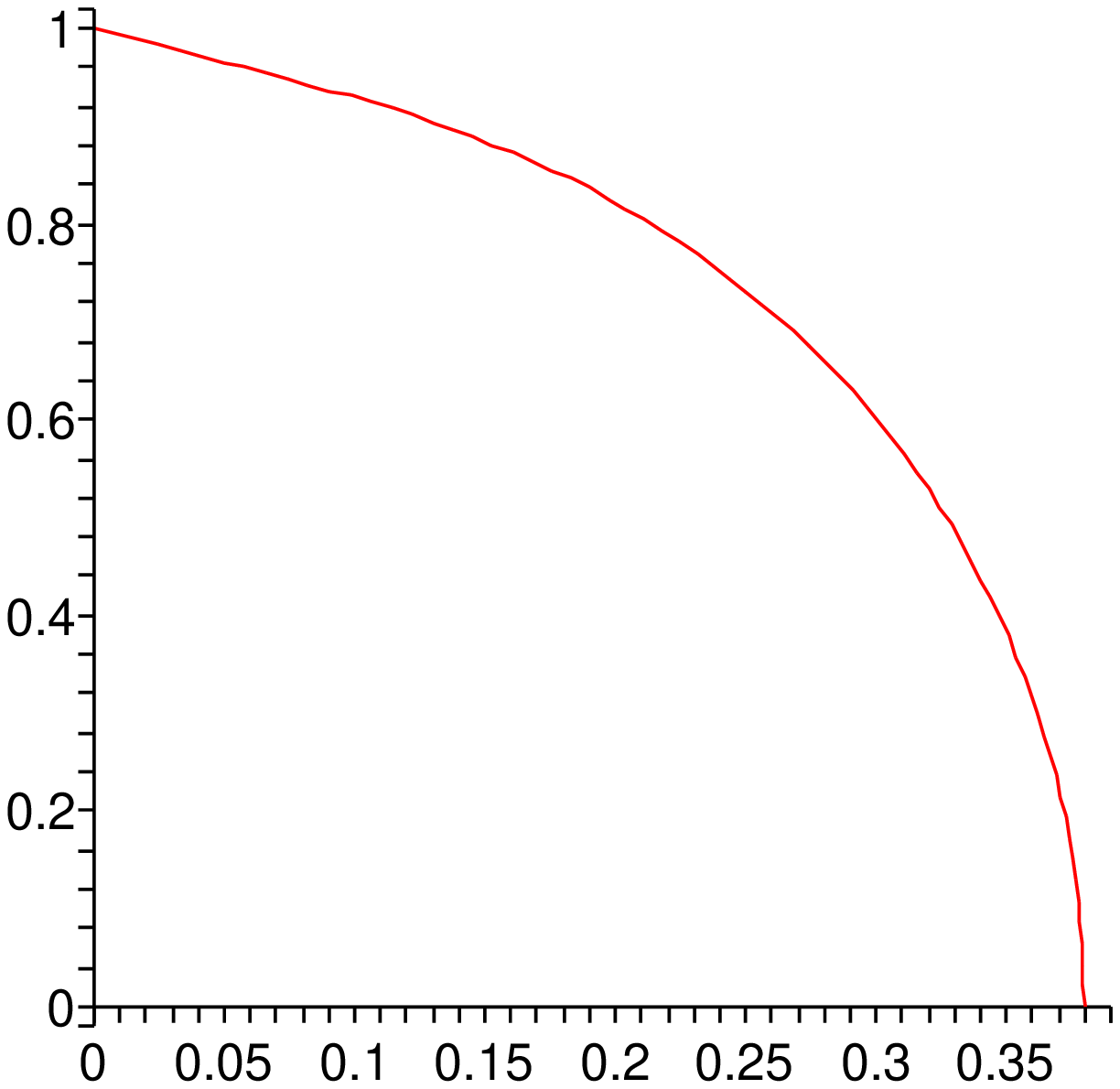,width=7 cm,height=6cm}
\hskip .5cm \epsfig{file=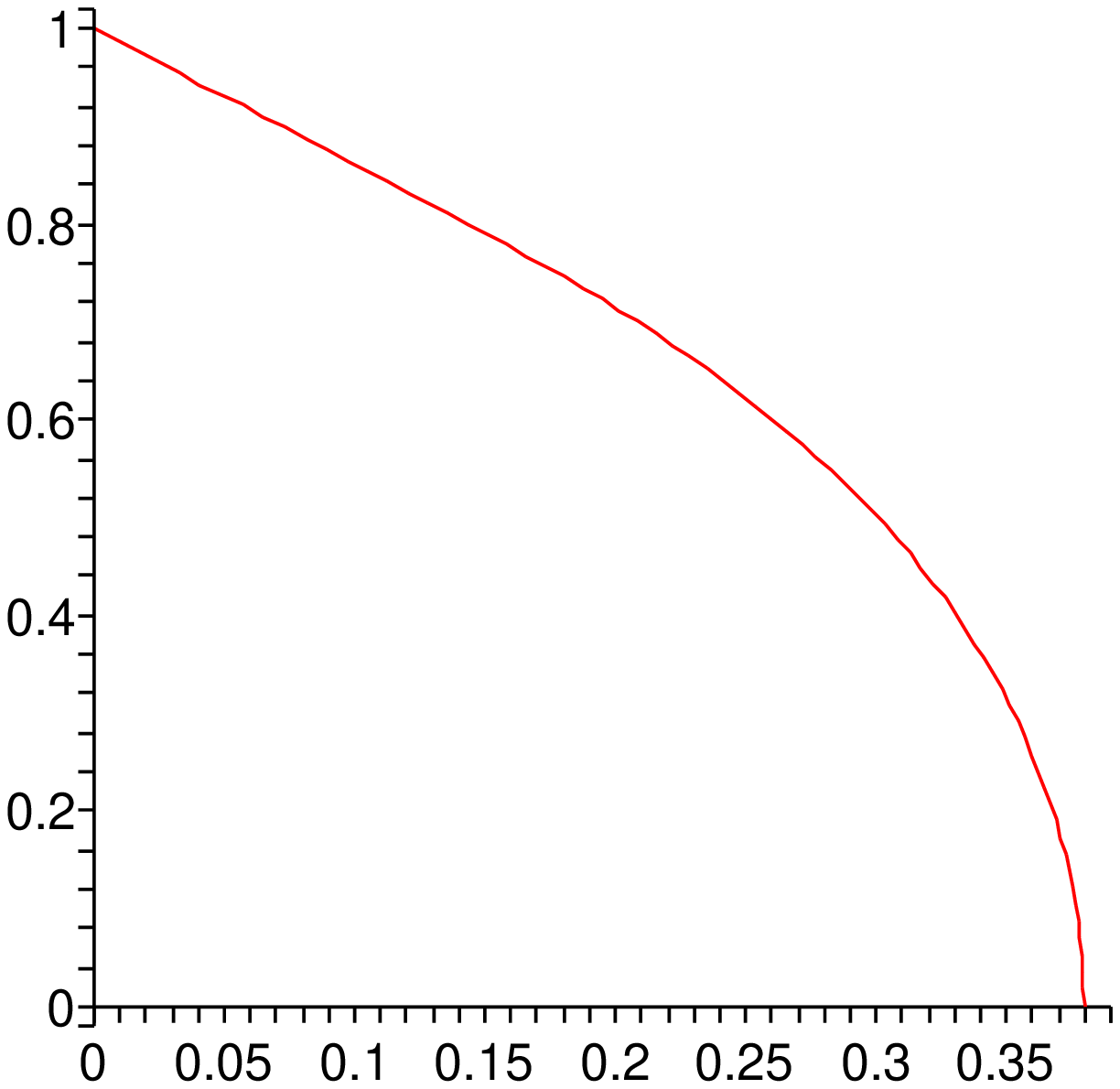,width=7 cm,height=6cm} }
\caption{The Hagedorn temperature $T_H$ as function of $\Omega$ in
the two cases $(\Omega_1,\Omega_2,\Omega_3)=(\Omega,\Omega,0)$,
displayed on the left, and
$(\Omega_1,\Omega_2,\Omega_3)=(\Omega,\Omega,\Omega)$, displayed
on the right.\label{figTO23}}
\begin{picture}(0,0)(0,0)
\put(5,207){\Large $\Omega$} \put(200,72){\Large $T$}
\put(222,207){\Large $\Omega$} \put(417,72){\Large $T$}
\end{picture}
\end{figure}

For small chemical potentials it is straightforward to compute
that the Hagedorn temperature is
\begin{equation}
\label{Thag} \begin{array}{c} \ds T_H (\Omega_1,\Omega_2,\Omega_3)
= \frac{1}{\beta_0} + p_1 \sum_{i=1}^3 \Omega_i^2 + p_2 \sum_{i<j}
\Omega_i^2 \Omega_j^2  + p_3 \sum_{i=1}^3 \Omega_i^4  + \CO
(\Omega_i^6 ) \\[5mm] \ds
 \beta_0 = -\log(7-4\sqrt{3}) \spa p_1
= - \frac{1}{6\sqrt{3}} \spa
p_2=\beta_0\frac{(18-5\sqrt{3})}{1296} \spa
p_3=\beta_0\frac{(18-11\sqrt{3})}{2592}
\end{array}
\end{equation}
In Figure \ref{figTO1} and Figure \ref{figTO23} we have displayed
$T_H$ as a function of $\Omega$ for the three particular cases
given by $(\Omega_1,\Omega_2,\Omega_3)=(\Omega,0,0)$,
$(\Omega_1,\Omega_2,\Omega_3)=(\Omega,\Omega,0)$ and
$(\Omega_1,\Omega_2,\Omega_3)=(\Omega,\Omega,\Omega)$. As we shall
see in the following those three special cases are highly relevant
for this paper. Note that if we define $R$ as being the charge
related to the chemical potential $\Omega$ we have that $R=R_1$,
$R=R_1+R_2$ and $R=R_1+R_2+R_3$ corresponds to the three cases,
respectively.

For the case $(\Omega_1,\Omega_2,\Omega_3)=(\Omega,0,0)$ depicted
in Figure \ref{figTO1} we see that the behavior near the critical
point $(T,\Omega)=(0,1)$ is
\begin{equation}
T_H(\Omega) = - \frac{1}{\log (1-\Omega)} \left[ 1 - \frac{\log (
-\log (1-\Omega) )}{\log (1-\Omega)} + \cdots \right]
\end{equation}
for $1 - \Omega \ll 1$. Thus, the slope of the Hagedorn curve in
the $(T,\Omega)$ diagram is zero in the critical point
$(T,\Omega)=(0,1)$, as is also clear from Figure \ref{figTO1}.

For the case $(\Omega_1,\Omega_2,\Omega_3)=(\Omega,\Omega,0)$
depicted in the left part of Figure \ref{figTO23} we have instead
that the behavior near the critical point $(T,\Omega)=(0,1)$ is
\begin{equation}
\label{thoII} T_H(\Omega) = \frac{1-\Omega}{\log 2} \left[ 1 -
\frac{2}{\log 2} e^{-\frac{1}{2} \log 2 / (1-\Omega) } + \CO (
e^{-\log 2 /(1-\Omega) } ) \right]
\end{equation}
for $1-\Omega \ll 1$. We see from this that the slope of the
Hagedorn curve at the critical point $(T,\Omega)=(0,1)$ is $-\log
2$.

Finally for the case
$(\Omega_1,\Omega_2,\Omega_3)=(\Omega,\Omega,\Omega)$ depicted in
the right part of Figure \ref{figTO23} we have that the behavior
near the critical point $(T,\Omega)=(0,1)$ is
\begin{equation}
\label{thoIII} T_H(\Omega) = \frac{1-\Omega}{\log 4} \left[ 1 -
\frac{6}{\log 4 } e^{-\log 4 / (1-\Omega)} + \CO ( e^{-2\log 4 /
(1-\Omega)} ) \right]
\end{equation}
for $1-\Omega \ll 1$. We see from this that the slope of the
Hagedorn curve at the critical point $(T,\Omega)=(0,1)$ is $-\log
4$.

\subsection{Decoupling for near-critical chemical potentials}
\label{sec:nearcrit}

We now turn to examine what happens when the chemical potentials
are near-critical, i.e. when one or more of the chemical
potentials $\Omega_i$ are close to 1. From Figure \ref{figTO1} and
Figure \ref{figTO23} we see that to zoom in to a region where the
chemical potentials are near-critical we also need to send the
temperature to zero. For $x \rightarrow 0$ it is clear that
$z(x,y_i) \rightarrow 0$ unless we send one or more of the $y_i$
to infinity (we restrict ourselves here to positive chemical
potentials without loss of generality). Write now $y_i =
y^{\alpha_i}$ where $\alpha_i$, $i=1,2,3$, are numbers. Assume
without loss of generality $0 \leq \alpha_3 \leq \alpha_2 \leq
\alpha_1 = 1$. From Eq.~\eqref{lettpart} we see then that we
should take the limit
\begin{equation}
\label{xylim} x \rightarrow 0 \spa xy = \mbox{fixed}
\end{equation}
One can now see that we get three different limits depending on if
one, two or three of the $\alpha_i$, $i=1,2,3$, are equal to one. It
is easy to see that this corresponds to sending either one, two or
three of the $\Omega_i$, $i=1,2,3$, to $1$ as $T \rightarrow 0$. We
can therefore restrict ourselves in the following to the three cases
$(\alpha_1,\alpha_2,\alpha_3) \in \{ (1,0,0),(1,1,0),(1,1,1) \}$.

Writing $y = \exp (\beta \Omega )$ we see that the limit
\eqref{xylim} means that $T \rightarrow 0$ and $\Omega \rightarrow
1$ such that $T/(1-\Omega)$ is fixed. In fact, it is useful to
define
\begin{equation}
\label{tildeT} \tilde{T} \equiv \frac{T}{1-\Omega} \spa \tilde{x}
\equiv xy \spa \tilde{x} = \exp( - 1 / \tilde{T})
\end{equation}
As we shall see, $\tilde{T}$ can be thought of as a temperature in
the decoupled sector after taking the limit \eqref{xylim}. The
R-charge that corresponds to the chemical potential $\Omega$ is $R
= \sum_{i=1}^3 \alpha_i R_i $. With this, we have $y_1^{R_1}
y_2^{R_2} y_3^{R_3} = y^R$.

\subsubsection*{Case I: $R=R_1$. The half-BPS sector}

We take $(\alpha_1,\alpha_2,\alpha_3)=(1,0,0)$ and hence $R =
R_1$. From the letter partition function \eqref{lettpart} we see
that in the limit \eqref{xylim} we have
\begin{equation}
\label{zcI} z(x,y_i) = xy = \tilde{x}
\end{equation}
up to corrections of order $x$. Therefore, we see that the set of
possible letters reduces to just the single letter $Z$, which is
the complex scalar in $\CN=4$ SYM with weight $(1,0,0)$. The
multi-trace operators in this sector are of the form
\begin{equation}
\label{halfBPSop} \tr ( Z^{L_1} ) \tr ( Z^{L_2} )  \cdots \tr (
Z^{L_k} )
\end{equation}
Thus, the limit we are considering corresponds to being in the
well-known half-BPS sector of $\CN=4$ SYM spanned by operators of
the form \eqref{halfBPSop}. All the operators of the form
\eqref{halfBPSop} are chiral primaries of $\CN=4$ SYM and preserve
at least half of the supersymmetries. By considering the partition
function \eqref{fullZN} for any $N$ we see that the partition
function of $U(N)$ $\CN=4$ SYM on $\R \times S^3$ reduces to the
one of the half-BPS sector given by \eqref{halfBPSop}. The limit
thus reduces the $\CN=4$ SYM to the quantum mechanical theory with
\eqref{halfBPSop} as the states in the Hilbert space. This was
previously discussed in \cite{Yamada:2005um}.

If we consider the thermodynamics of the half-BPS sector
\eqref{halfBPSop} for large $N$ it is easy to see from \eqref{zcI}
that we never reach the Hagedorn singularity: $\tilde{T}$ can be
arbitrarily large.

We note here that the half-BPS sector \eqref{halfBPSop} is
interesting for various reasons; it contains the states dual to
the vacuum of the maximally supersymmetric pp-wave
\cite{Blau:2001ne,Berenstein:2002jq}, to $AdS_5 \times S^5$
\cite{Aharony:1999ti}, and to giant gravitons in $AdS_5 \times
S^5$ \cite{McGreevy:2000cw}, and a correspondence between states
in this sector and half-BPS backgrounds of type IIB string theory
has been found in \cite{Lin:2004nb}.

\subsubsection*{Case II: $R=R_1+R_2$. The $SU(2)$ sector}

For this case we take $(\alpha_1,\alpha_2,\alpha_3)=(1,1,0)$ so
that $R = R_1+R_2$ and
$(\Omega_1,\Omega_2,\Omega_3)=(\Omega,\Omega,0)$. Taking the limit
\eqref{xylim} the letter partition function \eqref{lettpart} now
becomes
\begin{equation}
\label{zcII} z(x,y_i) = 2xy = 2\tilde{x}
\end{equation}
up to corrections of order $x$. In this case, the set of possible
letters reduces to the two complex scalars $Z$ and $X$ with weights
$(1,0,0)$ and $(0,1,0)$, respectively. This is due to the fact that
these two letters are the only letters for which the conformal
dimension is equal to the eigenvalue of $R=R_1+R_2$. For all other
letters the conformal dimension is greater than the eigenvalue of
$R$. Thus, the set of multi-operators consist of all operators of
the form
\begin{equation}
\label{su2op} \tr( A^{(1)}_1 A^{(1)}_2 \cdots A^{(1)}_{L_1} )\tr(
A^{(2)}_1 A^{(2)}_2 \cdots A^{(2)}_{L_2} ) \cdots \tr( A^{(k)}_1
A^{(k)}_2 \cdots A^{(k)}_{L_k} ) \spa A^{(i)}_{j} =Z,X
\end{equation}
From \eqref{fullZN} we see that the partition function for free
$U(N)$ $\CN=4$ SYM on $\R \times S^3$ in the limit \eqref{xylim}
is
\begin{equation}
\label{fullZII} Z(x,y_i) = \int [dU] \exp \left[ \sum_{k=1}^\infty
\frac{2\tilde{x}^k}{k} \tr ( U^k ) \tr ( (U^\dagger)^k ) \right]
\end{equation}
As for the half-BPS sector we see that $\CN=4$ SYM in the limit
\eqref{xylim} is reduced to a quantum mechanical theory, with the
multi-trace operators \eqref{su2op} as the Hilbert-space. It is
not hard to see that precisely the fact that $x\rightarrow 0$
means that the more covariant derivatives an operator has the more
decoupled it becomes. Thus, we remove all the modes coming from
having a field theory on a space, i.e.~in this case the
Kaluza-Klein modes on $S^3$. In this sense we lose the locality of
the field theory and the system becomes instead quantum
mechanical.

In Appendix \ref{app:osc} we take the limit \eqref{xylim} in the
oscillator representation of $\CN=4$ SYM. This is an alternative
way of showing that we get the $SU(2)$ sector in the limit
\eqref{xylim}.

For $N=\infty$ it is easy to see from \eqref{zcII} that we have a
Hagedorn singularity for $\tilde{x}=\frac{1}{2}$, which
corresponds to
\begin{equation}
\label{hagII} \tilde{T}_H = \frac{T_H(\Omega)}{1-\Omega} =
\frac{1}{\log 2}
\end{equation}
We note that this precisely corresponds to the leading part of
\eqref{thoII}. Indeed, viewing the limit \eqref{xylim} as zooming
into the region $T \ll 1$ and $1-\Omega \ll 1$ we see that
corresponds to the linear slope of the Hagedorn curve near the
critical point $(T,\Omega)=(0,1)$ in the left part of Figure
\ref{figTO23}.

In conclusion, we see that the $SU(2)$ sector captures the leading
features of $\CN=4$ SYM on $\R \times S^3$ near the critical point
$(T,\Omega)=(0,1)$. We also see that despite the fact that $\CN=4$
SYM reduces from a field theory to a quantum mechanical theory we
keep the interesting physics such as the Hagedorn transition for
large $N$. Finally we note that using the partition function
\eqref{fullZII} when $T \ll 1$ and $1-\Omega \ll 1$ instead of the
full partition function \eqref{fullZN} for free $\CN=4$ SYM on $\R
\times S^3$ is a very good approximation. Indeed, if $\Omega=0.99$,
the correction on the Hagedorn temperature is of order $10^{-15}$.

\subsubsection*{Case III: $R=R_1+R_2+R_3$. The $SU(2|3)$ sector}

This case has $(\alpha_1,\alpha_2,\alpha_3)=(1,1,1)$ and hence
$R=R_1+R_2+R_3$. Taking the limit \eqref{xylim} the letter
partition function \eqref{lettpart} reduces to
\begin{equation}
\label{zcIII} z(x,y_i) = 3xy + 2(xy)^{\frac{3}{2}} = 3\tilde{x} +
2\tilde{x}^{\frac{3}{2}}
\end{equation}
up to corrections of order $x$. Thus, the set of possible letters
reduces to the three complex scalars $Z$, $X$ and $W$ with weights
$(1,0,0)$, $(0,1,0)$ and $(0,0,1)$, respectively, and two complex
fermions $\chi_{1}$ and $\chi_2$ both of weight
$(\frac{1}{2},\frac{1}{2},\frac{1}{2})$.%
\footnote{Note here that we started with 16 real fermionic
components. Picking out a particular weight then leaves us with
two real fermionic components, corresponding to the
$2\tilde{x}^{3/2}$ term in the partition function. This can also
be seen as two complex fermions $\chi_{1}$ and $\chi_2$ in the
sense that their complex conjugates are not present in this
sector, just as the complex conjugates of the three complex
scalars $Z$, $X$ and $W$ are not present in this sector.}
This is precisely the $SU(2|3)$ sector of $\CN=4$ SYM as defined
in \cite{Beisert:2003jj,Beisert:2003ys}. In Appendix \ref{app:osc}
we have shown this using the oscillator representation of $\CN=4$
SYM. In this way we show directly that we obtain the $SU(2|3)$
sector as it is defined in \cite{Beisert:2003jj} in terms of the
oscillator representation of $\CN=4$ SYM.

The Hilbert space of the $SU(2|3)$ sector consists of the
multi-trace operators
\begin{equation}
\label{su23op} \begin{array}{c} \ds \tr( A^{(1)}_1 A^{(1)}_2
\cdots A^{(1)}_{L_1} )\tr( A^{(2)}_1 A^{(2)}_2 \cdots
A^{(2)}_{L_2} ) \cdots \tr( A^{(k)}_1 A^{(k)}_2 \cdots
A^{(k)}_{L_k} ) \\[2mm]
\ds A^{(i)}_{j} =Z,X,W,\chi_1,\chi_2
\end{array}
\end{equation}
From \eqref{fullZN} we see that the partition function for free
$U(N)$ $\CN=4$ SYM on $\R \times S^3$ in the limit \eqref{xylim}
is
\begin{equation}
Z(x,y_i) = \int [dU] \exp \left[ \sum_{k=1}^\infty
\frac{3\tilde{x}^k+  2(-1)^{k+1} \tilde{x}^{\frac{3}{2}k}}{k} \tr
( U^k ) \tr ( (U^\dagger)^k ) \right]
\end{equation}
For $N=\infty$ we see from \eqref{zcIII} that the Hagedorn
singularity occurs at
\begin{equation}
\label{hagIII}
\tilde{T}_H =\frac{T_H(\Omega)}{1-\Omega} =
\frac{1}{\log 4}
\end{equation}

\subsection{Above the Hagedorn temperature}
\label{sec:aboveTH}

In this section we consider the behavior of free $\CN=4$ SYM on
$\R \times S^3$ above the Hagedorn temperature, following
\cite{Aharony:2003sx}.%
\footnote{See also \cite{Liu:2004vy}.} Since the $N=\infty$
partition function is singular at the Hagedorn temperature we
should instead use the exact partition function \eqref{fullZN}
which takes non-trivial relations between multi-trace operators
into account. Now, the eigenvalues of the $U(N)$ group element $U$
are elements $e^{i\theta}$ on the unit circle. For large $N$ these
eigenvalues become a continuous distribution and we write
$\rho(\theta)$ for the density of eigenvalues at the angle
$\theta$ normalized such that $\int_{-\pi}^\pi d\theta
\rho(\theta) = 1$. Using this, we find from \eqref{fullZN} the
effective action for the eigenvalues \cite{Aharony:2003sx}
\begin{equation}
\label{effacteig}
I = N^2 \sum_{n=1}^\infty |\rho_n|^2 a_n \spa
a_n (x,y_i) = \frac{1 - z(\omega^{n+1} x^n, y_i^n)}{n}
\end{equation}
with $\rho_n = \int_{-\pi}^\pi d\theta \cos(n\theta) \rho(\theta)
$. To find the correct eigenvalue distribution we should minimize
$I$. For temperatures below the Hagedorn temperature we have that
$a_n > 0$ and hence the minimum distribution of eigenvalues is the
uniform distribution. This is easily seen to give the $N=\infty$
partition function \eqref{multinfty} \cite{Aharony:2003sx}.

When we reach the Hagedorn temperature we have that $a_1 = 0$, and
this means that the minimum of $I$ appears for a non-uniform
distribution of the eigenvalues when we are above the Hagedorn
temperature. Using the same procedure as in \cite{Aharony:2003sx}
we determine the behavior of the free energy near the transition
as a perturbative expansion in $\Delta T \equiv T-T_H(\Omega_i)$
when we are slightly above the Hagedorn temperature. Following
\cite{Aharony:2003sx}, the expression for the partition function
can be written as
\begin{equation}
-\frac{\log Z}{N^2} = -\frac{\epsilon^2}{4} -\frac{\epsilon^3}{3}-
\epsilon^4 \left(\frac{3}{8} - \frac{1}{4}\sum_{n=2}^{\infty}
\frac{n(n^2-1)z(x^n,y_i^n)}{1-z(x^n,y_i^n)}\right)+{\cal{O}}(\epsilon^5)
\label{above}
\end{equation}
where $\epsilon= \cos^2(\theta_0/2)$, the angle $\theta_0$ is
defined by $\sin^2 (\theta_0/2) = 1 - \sqrt{1-1/z(x,y_i)}$ and
$z(x,y_i)$ is given in Eq.~\eqref{lettpart}. The Gibbs free energy
$F=F(T,\Omega_i)$ slightly above the Hagedorn temperature is then
given by
\begin{equation}
\frac{F}{N^2}= -\frac{1}{4} \left. \frac{\partial
z(x,y_i)}{\partial T} \right|_{T=T_H}T_H \Delta T
-\frac{1}{3}\left( \left. \frac{ \partial z(x,y_i)}{\partial
T}\right|_{T=T_H} \right)^{3/2}T_H \Delta T ^{3/2} +{\cal{O}}
(\Delta T^{2}) \label{freeabove}
\end{equation}
with $\Delta T \equiv T-T_H(\Omega_i) \geq 0$. Using
\eqref{freeabove} with \eqref{Thag} we get the explicit expansion
\begin{equation}
\begin{array}{rcl}
\ds \frac{F}{N^2}&=& \ds
-\beta_0\frac{3}{8}\left(1-\beta_0\frac{(2\sqrt{3}+\beta_0)}{36}\sum_{i=1}^3\Omega_i^2
+\CO(\Omega_i^4) \right)\Delta T \\[4mm] && \ds -\beta_0^2\sqrt{\frac{3}{8}}\left(1-\beta_0\frac{(4+\sqrt{3}\,\beta_0)}{24 \sqrt{3}}
\sum_{i=1}^3\Omega_i^2+\CO(\Omega_i^4)\right)\Delta T ^{3/2}+
{\cal{O}}(\Delta T^{2})
\end{array}
\label{feabove}
\end{equation}
for $0 \leq \Delta T \ll 1$. When the chemical potentials are set
to zero we recover the result of \cite{Aharony:2003sx}.

Note from the above that while $F/N^2$ in the large $N$ limit is
finite for temperatures above the Hagedorn temperature, it is zero
for temperatures below the Hagedorn temperature. Thus, we can regard
$F/N^2$ as an order parameter for the Hagedorn phase transition.
Since the derivative of the free energy is discontinuous at the
Hagedorn temperature we see that free $U(N)$ $\CN=4$ SYM on
$\R\times S^3$ has a first order phase transition at the Hagedorn
temperature \cite{Aharony:2003sx}.

We now turn to the behavior of the free energy slightly above the
Hagedorn temperature in the case of near-critical chemical
potential. We examine the two cases corresponding to
$(\Omega_1,\Omega_2,\Omega_3)=(\Omega,\Omega,0)$ and
$(\Omega_1,\Omega_2,\Omega_3)=(\Omega,\Omega,\Omega)$ sending
$\Omega \rightarrow 1$ by taking the limit \eqref{xylim} described
in Section \ref{sec:nearcrit}. In this limit we get a rescaled
temperature $\tilde{T} = T/(1-\Omega)$ as defined in
\eqref{tildeT}. From this we see that we naturally get a rescaled
free energy $\tilde{F} = - \tilde{T} \log Z = F / (1-\Omega)$
where $F$ is the Gibbs free energy. In the limit $\Omega
\rightarrow 1$ with $\tilde{T}$ fixed, we get that
$\tilde{F}=\tilde{F} ( \tilde{T} )$, i.e. the rescaled free energy
depends only on $\tilde{T}$.

Considering the case
$(\Omega_1,\Omega_2,\Omega_3)=(\Omega,\Omega,0)$ we have from
Section \ref{sec:nearcrit} that free $\CN=4$ SYM decouples to the
$SU(2)$ sector \eqref{su2op} in the limit $\Omega\rightarrow 1$
with $\tilde{T}$ fixed. From \eqref{hagII} we have that the
Hagedorn temperature is $\tilde{T}_H = 1/\log 2$. Using
\eqref{freeabove}, it is straightforward to show that the free
energy slightly above the Hagedorn temperature is
\begin{equation}
\frac{\tilde{F}}{N^2}=-\frac{\log{2}}{4}(\tilde{T}-\tilde{T}_H)
-\frac{(\log{2})^{2}}{3}(\tilde{T}-\tilde{T}_H)^{3/2} +
{\cal{O}}((\tilde{T}-\tilde{T}_H)^{2}) \label{feabove22}
\end{equation}
for $0 \leq \tilde{T} - \tilde{T}_H \ll 1$. One can either derive
this using the full letter partition function \eqref{lettpart} and
then take the limit $\Omega\rightarrow 1$ with $\tilde{T}$ fixed,
or alternatively derive it directly using the letter partition
function \eqref{zcII} for the $SU(2)$ sector.

Similarly we can proceed in the case
$(\Omega_1,\Omega_2,\Omega_3)=(\Omega,\Omega,\Omega)$ where we
have from Section \ref{sec:nearcrit} that free $\CN=4$ SYM
decouples to the $SU(2|3)$ sector \eqref{su23op} in the limit
$\Omega\rightarrow 1$ with $\tilde{T}$ fixed. We know from
Eq.~\eqref{hagIII} that the Hagedorn temperature is
$\tilde{T}_H=1/\log 4$ and using \eqref{freeabove} we have that
the free energy slightly above the Hagedorn temperature is
\begin{equation}
\frac{\tilde{F}}{N^2}=-\frac{9\log 2}{16}(\tilde{T}-\tilde{T}_H)
-\frac{9(\log{2})^2}{4\sqrt{2}}(\tilde{T}-\tilde{T}_H)^{3/2} +
{\cal{O}}((\tilde{T}-\tilde{T}_H)^{2}) \label{feabove23}
\end{equation}
for $0 \leq \tilde{T} - \tilde{T}_H \ll 1$. Again, as in the
$SU(2)$ sector, this result can be found in two different ways
corresponding to either starting from the letter partition
function \eqref{lettpart} and then take the limit on the final
result, or starting with the $SU(2|3)$ letter partition function
\eqref{zcIII}.

\subsubsection*{High temperatures}

If we consider instead the high temperature regime the eigenvalue
distribution becomes almost like a delta-function
\cite{Aharony:2003sx}. Therefore, $\rho_n = 1$ and we get that $I
= N^2 \sum_{n=1}^\infty a_n$. If we consider a high-temperature
limit with the chemical potentials being fixed, we get the Gibbs
free energy
\begin{equation}
F = - \frac{\pi^2}{6} V ( S^3 ) N^2 T^4 + \CO (T^3)
\end{equation}
This is precisely the free energy of free $\CN=4$ SYM, i.e. it is
the result that one would get from $N^2$ times the free energy of
$U(1)$ $\CN=4$ SYM. Thus while free $\CN=4$ SYM on $\R \times S^3$
behaves as a confined theory for low temperature, it behaves as a
deconfined theory at high temperatures
\cite{Sundborg:1999ue,Aharony:2003sx}.

If we instead consider the case in which $\Omega_i/T$ does not go to
zero for $T \rightarrow \infty$ for at least one of the chemical
potentials, we get the free energy
\begin{equation}
F=-V_{S^3}N^2\left[\frac{\pi^2}{6}T^4+\frac{1}{4}T^{2}\sum_{i=1}^3\Omega_i^{2}
-\frac{1}{32\pi^2}\left(\sum_{i=1}^3\Omega_i^4-2\sum_{i<j}\Omega_i^2\Omega_j^2\right)\right]
+{\cal O}(T^3) \label{tfree}
\end{equation}
This is the same result as in~\cite{Cvetic:1999rb,Harmark:1999xt}
where the free energy is computed as $N^2$ times the free energy
of free $U(1)$ $\CN=4$ SYM. Note that the regularization procedure
for obtaining \eqref{tfree} is the same as in
\cite{Cvetic:1999rb,Harmark:1999xt}.

\section{Quantum mechanical sectors for near-critical chemical potential}
\label{sec:QMsec}

In Section \ref{sec:nearcrit} we saw for free $\CN=4$ SYM on $\R
\times S^3$ that regions with small temperature and near-critical
chemical potential are very interesting since the free $\CN=4$ SYM
effectively reduces to free quantum mechanical systems in such
regions. In this section we continue to examine these quantum
mechanical sectors of thermal $\CN=4$ SYM on $\R \times S^3$ but now
in the full interacting theory. We show in the following that the
interacting $\CN=4$ SYM on $\R \times S^3$ reduces to well-defined
interacting quantum mechanical systems in such regions of small
temperature and near-critical chemical potential.

Consider the partition function
\begin{equation}
\label{PF1} Z(\beta,\Omega) = {\tr}_M \left( e^{-\beta D + \beta
\Omega R} \right)
\end{equation}
Here $D$ is the dilatation operator of $\CN=4$ on $\R \times S^3$
which for weak coupling $\lambda \ll 1$ can be expanded as
\cite{Beisert:2003tq,Beisert:2004ry}
\begin{equation}
\label{fullD} D = D_0 + \sum_{n=2}^\infty \lambda^{n/2} D_{n}
\end{equation}
where we define for convenience the 't Hooft coupling as
\begin{equation}
\label{coupling}
\lambda = \frac{\gym^2 N}{4\pi^2}
\end{equation}
Furthermore, $R$ is a linear combination of the three R-charges
$R_1$, $R_2$ and $R_3$, with $\Omega$ as the corresponding
chemical potential. We restrict in the following to the three
cases $R=R_1$, $R=R_1+R_2$ and $R=R_1+R_2+R_3$. Clearly we have
that $D_0 \geq R$ for the three choices of $R$.

We can rewrite the partition function \eqref{PF1} as follows
\begin{equation}
\label{PF2} Z(\beta,\Omega) = {\tr}_M \left( \exp \left[ - \beta
(D_0-R) -\beta (1-\Omega) R - \beta \lambda D_2 - \beta
\sum_{n=3}^\infty \lambda^{n/2} D_{n} \right] \right)
\end{equation}
Consider the region
\begin{equation}
\label{thereg} T \ll 1 \spa 1 - \Omega \ll 1 \spa \lambda \ll 1
\end{equation}
We now argue that one can neglect all states with $D_0-R > 0$ in
the partition function \eqref{PF2}. First we observe that since
$\beta \gg 1$ and $D_0 - R$ is a non-negative integer the states
with $D_0-R > 0$ would have an exceedingly small weight factor.
However, one should also ensure then that the $D_0 = R$ states
does not have an equally small weight factor. This is precisely
ensured by having $1-\Omega$ and $\lambda \ll 1$. We can therefore
write the partition function \eqref{PF1} in the region
\eqref{thereg} as
\begin{equation}
\label{PF3} Z(\beta,\Omega) = {\tr}_\CH \left( \exp \left[ -\beta
(1-\Omega) D_0 - \beta \lambda D_2 - \beta \sum_{n=3}^\infty
\lambda^{n/2} D_{n} \right] \right)
\end{equation}
with
\begin{equation}
\CH = \left\{ \alpha \in M \Big| (D_0-R) \alpha = 0 \right\}
\end{equation}
i.e. we have restricted the trace to be only over states with
$D_0=R$. Comparing this to Section \ref{sec:nearcrit}, we see that
restricting to states in $\CH$ corresponds to the reduction of
$\CN=4$ SYM on $\R \times S^3$ found in the free theory. Defining
\begin{equation}
\tilde{\beta} = \beta (1-\Omega) \spa \tilde{\lambda} =
\frac{\lambda}{1-\Omega}
\end{equation}
we can write \eqref{PF3} as
\begin{equation}
\label{PF4} Z(\tilde{\beta}) = {\tr}_\CH \left( e^{-\tilde{\beta}
H } \right)
\end{equation}
with $H$ being the Hamiltonian
\begin{equation}
\label{Hint} H =D_0 + \tilde{\lambda} D_2 + \tilde{\lambda}
\sqrt{\lambda} \sum_{n=0}^\infty \lambda^{n/2} D_{n+3}
\end{equation}
Considering the three cases $R=R_1$, $R=R_1+R_2$ and
$R=R_1+R_2+R_3$ we have from Section \ref{sec:nearcrit} that $\CH$
in those three cases corresponds to the half-BPS-sector given by
\eqref{halfBPSop}, the $SU(2)$ sector given by \eqref{su2op} and
the $SU(2|3)$ sector given by \eqref{su23op}. We have thus shown
that interacting $\CN=4$ SYM on $\R \times S^3$ reduces to those
sectors in the region \eqref{thereg} with the Hamiltonian given by
\eqref{Hint}.

Note that we have not assumed anything about $N$, thus the above
considerations work equally well for finite $N$ and in the large
$N$ limit. If we assume $N=\infty$, we can ignore the non-trivial
relations between multi-trace operators and work instead with
single-trace operators. We can then think of the Hamiltonian
\eqref{Hint} as the Hamiltonian of a periodic one-dimensional
spin-chain. Below we consider the three possible cases and
identify the spin-chain models.

\subsubsection*{Case I: $R=R_1$. The half-BPS sector}

For $R=R_1$ the interacting thermal $\CN=4$ SYM on $\R \times S^3$
is reduced to the Hamiltonian \eqref{Hint} acting on the
multi-trace operators of the form \eqref{halfBPSop}. Since these
operators are chiral primaries of $\CN=4$ SYM all the interaction
terms are zero on these states, and hence the Hamiltonian
\eqref{Hint} is $H=D_0$ for this sector.

\subsubsection*{Case II: $R=R_1+R_2$. The $SU(2)$ sector}

With $R=R_1+R_2$ the interacting thermal $\CN=4$ SYM on $\R \times
S^3$ is reduced to a quantum mechanical theory with Hamiltonian
\eqref{Hint} acting on the $SU(2)$ sector of $\CN=4$ SYM on $\R
\times S^3$ which is spanned by operators of the form
\eqref{su2op}. Note that in the $SU(2)$ sector the half-integer
powers of $\lambda$ in \eqref{Hint} are not present and we have
instead a Hamiltonian of the form \cite{Beisert:2003tq}
\begin{equation}
\label{Hintsu2} H =D_0 + \tilde{\lambda} D_2 + \tilde{\lambda}
\lambda \sum_{n=0}^\infty \lambda^{n} D_{2n+4}
\end{equation}

For $N=\infty$ we can restrict ourselves to consider the
single-trace operators, since they are a well-defined subset of
the operators. In the $SU(2)$ sector the single-trace operators
are of the form
\begin{equation}
\label{singsu2} \tr( A_1 A_2 \cdots A_{L} ) \spa A_i \in \{ X,Z \}
\end{equation}
Such single-trace operators can be regarded as spin-chains. In
particular a single-trace of length $L$ corresponds to a periodic
spin chain of length $L$. For a chain of length $L$ the leading
interaction term $D_2$ in the Hamiltonian \eqref{Hintsu2} is given
by \cite{Minahan:2002ve,Beisert:2003tq}
\begin{equation}
\label{D2su2} D_2 = \frac{1}{2} \sum_{i=1}^L (I_{i,i+1} -
P_{i,i+1} )
\end{equation}
Here $P_{i,i+1}$ is the permutation operator and $I_{i,i+1}$ is
the identity operator acting on the letters at positions $i$ and
$i+1$. This term of the Hamiltonian \eqref{Hintsu2} corresponds
precisely to the Hamiltonian of the ferromagnetic $XXX_{1/2}$
Heisenberg spin chain reviewed in Appendix \ref{app:XXX}, where we
think of the letters $Z$ and $X$
as spin up and spin down.%
\footnote{Note that $J = - \tilde{\lambda}$ in comparing with the
Hamiltonian \eqref{HXXX}.} Some of the higher terms in
\eqref{Hintsu2} are known as well
\cite{Beisert:2003tq,Beisert:2004hm}, but as will be clear in the
following they will not play a role for our considerations since
they are much weaker coupled than the $D_2$ term. Finally we note
that there is considerable evidence that the Hamiltonian
\eqref{Hintsu2} is integrable
\cite{Minahan:2002ve,Beisert:2003tq,Beisert:2003yb,Beisert:2004hm}.

For $N=\infty$ we can thus conclude that the thermodynamics of
$\CN=4$ SYM on $\R \times S^3$ in the region \eqref{thereg} with
$R=R_1+R_2$ can be understood from the thermodynamics of the
$XXX_{1/2}$ Heisenberg spin chain.

\subsubsection*{Case III: $R=R_1+R_2+R_3$. The $SU(2|3)$ sector}

For the case $R=R_1+R_2+R_3$ the interacting thermal $\CN=4$ SYM on
$\R \times S^3$ in the region \eqref{thereg} reduces to a quantum
mechanical theory with Hamiltonian \eqref{Hint} acting on the
$SU(2|3)$ sector of $\CN=4$ SYM spanned by operators of the form
\eqref{su23op}.

When $N=\infty$ we can again restrict to the single-trace
operators which in this sectors are of the form
\begin{equation}
\label{singsu23} \tr( A_1 A_2 \cdots A_{L} ) \spa A_i \in \{
X,Z,W,\chi_1,\chi_2 \}
\end{equation}
Then a single-trace operator of length $L$ can be regarded as a
periodic spin-chain of length $L$. The leading interaction term
$D_2$ in the Hamiltonian \eqref{Hint} can then be written as
\cite{Beisert:2003jj,Beisert:2003ys}
\begin{equation}
\label{D2su23} D_2 = \frac{1}{2} \sum_{i=1}^L ( I_{i,i+1} -
\Pi_{i,i+1} )
\end{equation}
where $\Pi_{i,i+1}$ is the graded permutation operator which
permutes the fields at sites $i$ and $i+1$ picking up a minus sign
if the exchange involves two fermions.

In conclusion we have found that for $N=\infty$ the thermodynamics
of $\CN=4$ SYM on $\R \times S^3$ in the region \eqref{thereg}
with $R=R_1+R_2+R_3$ can be understood from the thermodynamics of
the $SU(2|3)$ spin chain with Hamiltonian \eqref{D2su23}.%
\footnote{Note that for this sector the spin-chain is dynamic
since it can change the length through the $D_3$ term
\cite{Beisert:2003ys}. However, we can ignore this higher-loop
effect here since we are mostly concerned with the one-loop
interaction which corresponds to the $D_2$ term \eqref{D2su23}.}

\section{Low temperature limit and the Heisenberg spin chain}
\label{sec:lowtemp}

In this section we consider what happens as we approach the
critical point $(T,\Omega)=(0,1)$ in the specific case of the
$SU(2)$ model, i.e. the case with $R=R_1+R_2$.

We saw in Section \ref{sec:QMsec} that the thermal partition
function of $\CN=4$ SYM on $\R \times S^3$ in the region
\eqref{thereg} with $R=R_1+R_2$ reduces to the partition function
\eqref{PF4} with the Hamiltonian \eqref{Hintsu2}. For $N=\infty$
we have that a single-trace of fixed length $L$ corresponds to
periodic spin-chain of length $L$ and the Hamiltonian
\eqref{Hintsu2} is a spin-chain Hamiltonian, with the leading
interaction term $D_2$ corresponding to an $XXX_{1/2}$ Heisenberg
spin chain Hamiltonian.

Consider now being in the region \eqref{thereg}. Take then the
zero temperature limit $T \rightarrow 0$ keeping $\lambda$ and
$\tilde{T}=T/(1-\Omega)$ fixed. In the $(T,\Omega)$ diagram
depicted in the left part of Figure \ref{figTO23} this corresponds
to moving towards the critical point $(T,\Omega)=(0,1)$ in a
straight line with slope $1/\tilde{T}$. In terms of the partition
function \eqref{PF4} and Hamiltonian \eqref{Hintsu2} we see that
this corresponds to fixing the temperature while increasing the
$\tilde{\lambda}$ coupling. Since we have that $\lambda \ll 1$ and
since $\tilde{\lambda}$ is growing towards infinity, we can ignore
the higher terms in \eqref{Hintsu2} and instead work with the
Hamiltonian
\begin{equation}
H = D_0 + \tilde{\lambda} D_2
\end{equation}
with $D_2$ given by \eqref{D2su2}. For a fixed length $L$ of the
chain (or for the single-trace operators) this is precisely the
ferromagnetic $XXX_{1/2}$ Heisenberg spin chain Hamiltonian (plus
a constant term). Therefore, we see that the approach to the
critical point $(T,\Omega)=(0,1)$ is governed completely by the
$XXX_{1/2}$ Heisenberg spin chain. Note that letting
$\tilde{\lambda}$ go to infinity does not spoil our approximations
of Section \ref{sec:QMsec} since we always have that $\beta \gg
\tilde{\beta}$.

Since we are keeping $\tilde{\beta}$ fixed we see from the weight
factor $e^{-\tilde{\beta} L -  \tilde{\beta}\tilde{\lambda} D_2}$
that it is reasonable to consider the limit for a chain of fixed
length since the coupling in front of $L$ is constant. The
remaining part of the weight factor is $e^{ -
\tilde{\beta}\tilde{\lambda} D_2}$ and thus we see that our limit
corresponds to taking the zero temperature limit of the
$XXX_{1/2}$ Heisenberg spin chain.

As reviewed in Appendix \ref{app:XXX} we have that the states with
the lowest energy of the ferromagnetic $XXX_{1/2}$ Heisenberg spin
chain are the zero eigenvalue states of $D_2$, which when written
as single-trace operators are of the form
\begin{equation}
\label{heivac} \tr \left( \mbox{sym} ( Z^{L-M} X^{M} ) \right)
\end{equation}
where 'sym' means total symmetrization. It is clear that any state
which is totally symmetrized has eigenvalue one under the
permutation operator, hence the eigenvalue of $D_2$ is zero on
such states. Since $0 \leq M \leq L$ we have $L+1$ different
vacuum states for a chain of length $L$. Now, since our limit
corresponds to taking the zero temperature limit of the
$XXX_{1/2}$ Heisenberg spin chain, and since the zero temperature
limit means that the states with lowest energy dominates, we can
conclude that we are driven towards the vacuum states
\eqref{heivac} as we approach the critical point
$(T,\Omega)=(0,1)$.

That we are driven towards the states \eqref{heivac} makes sense
also from another point of view, namely that \eqref{heivac}
corresponds to chiral primaries of $\CN=4$ SYM, and thus the zero
temperature limit that we are taking is driving us towards a $1/2$
BPS sector of $\CN=4$ SYM.

There is also another zero temperature limit which is natural to
consider. Start again in the region \eqref{thereg}. Let then $T
\rightarrow 0$ with $\Omega$ and $\lambda$ being fixed. In this
limit we have that the rescaled temperature $\tilde{T}$ decreases,
while the couplings $\tilde{\lambda}$ and $\lambda$ both are
fixed. This means that this limit corresponds to keeping the
Hamiltonian \eqref{Hintsu2} fixed while changing the temperature
$\tilde{T}$ of the decoupled theory. Thus, in this limit we are
moving towards the ground states of the quantum mechanical theory
given by the Hamiltonian \eqref{Hintsu2}. For $N=\infty$ we can
consider the single-trace operators of a fixed length. Then the
$D_0$ term in \eqref{Hintsu2} can be ignored and to leading order
(neglegting the $D_4$ term and higher terms) we have a zero
temperature limit of the ferromagnetic $XXX_{1/2}$ Heisenberg spin
chain. As for the previous limit considered above, this means we
are driven towards the ferromagnetic vacuum states \eqref{heivac},
which are chiral primaries of $\CN=4$ SYM.

We considered in the above two zero temperature limits of the
$SU(2)$ sector. It is not hard to see that we get similar results
for the corresponding zero temperature limits in the $SU(2|3)$
sector. In particular, we are driven towards the vacuum states of
the $SU(2|3)$ spin chain given by \eqref{D2su23} which are the
states that have zero eigenvalue for $D_2$. Moreover, these states
are chiral primaries of $\CN=4$ SYM.

\section{Decoupling limit to exact quantum mechanical Hamiltonian}
\label{sec:declim}

We show in the following that we can take decoupling limits of the
thermal interacting $\CN=4$ SYM on $\R \times S^3$ to a quantum
mechanical system which is described exactly by the one-loop
corrected Hamiltonian in that sector. For $N=\infty$ the
Hamiltonians are the ones of the well-known spin-chain models.

Consider the partition function \eqref{PF1} with the full
dilatation operator \eqref{fullD}. We consider here again the
cases $R=R_1$, $R=R_1+R_2$ and $R=R_1+R_2+R_3$. Consider then the
following decoupling limit
\begin{equation}
\label{declim} T \rightarrow 0 \spa \tilde{T} \equiv
\frac{T}{1-\Omega} \ \mbox{fixed} \spa \tilde{\lambda} \equiv
\frac{\lambda}{1-\Omega} \ \mbox{fixed}
\end{equation}
Clearly $\Omega \rightarrow 1$ and $\lambda \rightarrow 0$ in this
limit. From the partition function \eqref{PF1} it is clear that we
can ignore states with $D_0 > R$, and hence we only have states
with $D_0=R$. Applying the arguments of Section \ref{sec:QMsec} we
get that the limit \eqref{declim} of the full partition function
\eqref{PF1} reduces to the limit \eqref{declim} of the reduced
partition function \eqref{PF3}. Since $\lambda \rightarrow 0$ we
see that all the higher-loop terms drop out, and only the $D_0$
and $D_2$ terms remain. The limit \eqref{declim} of the partition
function \eqref{PF1} therefore gives the result
\begin{equation}
Z(\tilde{\beta}) = {\tr}_\CH \left( e^{-\tilde{\beta} H} \right)
\end{equation}
where $H$ is the Hamiltonian
\begin{equation}
\label{Hexact} H = D_0 + \tilde{\lambda} D_2
\end{equation}
Thus, thermal interacting $\CN=4$ SYM on $\R \times S^3$ in the
limit \eqref{declim} is described exactly by the Hamiltonian
\eqref{Hexact}. Note here that this is true for any $N$.
Furthermore, it is interesting to note that $\tilde{\lambda}$ can
take any value. One can thus end up with a strongly coupled $D_2$
term in the Hamiltonian as a good description of $\CN=4$ SYM, as
we in fact already saw in Section \ref{sec:lowtemp}.

For $N=\infty$, we get as above that we can think of the
single-trace operators as spin-chains. We thus have that the
thermodynamics of interacting $\CN=4$ SYM on $\R \times S^3$ in
the decoupling limit \eqref{declim} can be described exactly by a
spin-chain model with Hamiltonian \eqref{Hexact}.

If we consider the case $R=R_1+R_2$ for $N=\infty$ we see that the
thermodynamics of $\CN=4$ SYM on $\R \times S^3$ in the decoupling
limit \eqref{declim} can be described exactly by the ferromagnetic
$XXX_{1/2}$ Heisenberg spin chain (see Appendix \ref{app:XXX}).
This is easily seen from the Hamiltonian \eqref{Hexact} with $D_2$
given in \eqref{D2su2}. Written explicitly, we have that the full
partition function for $\CN=4$ SYM in the limit \eqref{declim} is
\begin{equation}
\log Z (\tilde{T}) = \sum_{n=1}^\infty \sum_{L=1}^\infty
\frac{1}{n} e^{-nL/\tilde{T}} Z^{(XXX)}_L (\tilde{T}/n)
\end{equation}
where $Z^{(XXX)}_L$ is the partition function for the
ferromagnetic $XXX_{1/2}$ Heisenberg spin chain of length $L$ with
Hamiltonian $\tilde{\lambda} D_2$.

Similarly, for the case $R=R_1+R_2+R_3$ we have that the
thermodynamics of $\CN=4$ SYM on $\R \times S^3$ in the decoupling
limit \eqref{declim} can be described exactly by the spin chain
model given by the $D_2$ term \eqref{D2su23}.

In conclusion we have found limits in which planar thermal $\CN=4$
SYM is described exactly by well-defined spin-chain models. The
spin-chain models involved are short-range and the coupling
$\tilde{\lambda}$ in front of the spin chain term $D_2$ in the
Hamiltonian can take any value.

\section{One-loop partition function}
\label{sec:oneloop}

In this section we consider the one-loop correction to the
partition function for $U(N)$ $\CN=4$ SYM on $\R \times S^3$ with
non-zero chemical potentials in planar limit $N=\infty$,
generalizing the procedure in~\cite{Spradlin:2004pp}. We use this
to find the one-loop correction to the Hagedorn temperature. We
consider subsequently the one-loop correction to the partition
function and Hagedorn temperature for the near-critical regions
where $\CN=4$ SYM reduces to the $SU(2)$ and $SU(2|3)$ sectors.

\subsubsection*{One-loop correction to partition function and Hagedorn temperature}

Consider the complete single-trace partition function ${Z}_{\rm
ST} = \tr ( x^D \prod_{i=1}^3 y_i^{R_i} )$ for $U(N)$ $\CN=4$ SYM
on $\R \times S^3$ in the planar limit. Up to the first order in
the 't Hooft coupling $\lambda$ the single-trace partition
function can be written as ${Z}_{\rm ST}={Z}_{\rm
ST}^{(0)}+\lambda {Z}_{\rm ST}^{(1)}+\CO(\lambda^2)$ where
${Z}_{\rm ST}^{(0)}$ is the zeroth order single-trace partition
function given in \eqref{ZST} and with the first-order
contribution given by
\begin{equation}
{Z}_{\rm ST}^{(1)}(x,y_i)=\log x \tr \left[ \prod_{i=1}^3
y_i^{R_i} x^{D_0} D_2 \right] \label{stpf}
\end{equation}
This follows from the expansion \eqref{fullD} of the dilatation
operator and from the fact that the R-charges commute with the
dilatation operator. Applying the arguments
of~\cite{Spradlin:2004pp} where it is used that one can refrase
\eqref{stpf} as a spin-chain partition function, we arrive at the
following expression for the one-loop single trace partition
function
\begin{equation}
\begin{array}{rcl} \ds
{Z}_{\rm ST}^{(1)}(x,y_i)&=& \ds \log x \sum_{L=1}^{\infty}
\sum_{\tiny \begin{array}{c} k=0 \\[-1mm] (k,L)=1 \end{array}}^{L-1}
\left(\frac{\langle
D_2(\omega^{L+1}x^{L},y_i^{L})\rangle}{1-z(\omega^{L+1}x^{L},y_i^{L})}\right.
\\[7mm] && \ds + \left. \delta_{L\neq 1}\langle
PD_2(\omega^{L-k+1}x^{L-k},y_i^{(L-k)},\omega^{k+1}x^{k},y_i^{k})\rangle
\right)
\end{array} \label{final}
\end{equation}
with
\begin{eqnarray}
& \ds \langle D_2(x,y_i)\rangle=\sum_{A_1,A_2 \in
\mathcal{A}}\prod_{i=1}^3x^{d(A_1)+d(A_2)}y_i^{R_i(A_1)+R_i(A_2)}\langle
A_1A_2|D_2|A_1A_2\rangle & \label{Oy}
\\
& \ds \langle PD_2(w,y_i,\bar{w},\bar{y}_i)\rangle=\sum_{A_1,A_2
\in
\mathcal{A}}\prod_{i=1}^3w^{d(A_1)}y_i^{R_i(A_1)}\bar{w}^{d(A_2)}\bar{y}_i^{R_i(A_2)}\langle
A_1A_2|D_2|A_2A_1\rangle & \label{POy}
\end{eqnarray}
Here $L$ can be seen as the length of the spin chain and $(k,L)=1$
means that $k$ and $L$ are relatively prime. We have also included
the fermion contribution. We note that Eq.~\eqref{final} is a
direct generalization of the result of~\cite{Spradlin:2004pp}.
From \eqref{final} it is in principle straightforward to compute
the one-loop correction \eqref{stpf}, once the two expectation
values \eqref{Oy} and \eqref{POy} are known. From this one gets
the corrected multi-trace partition function using the general
prescription in \eqref{multinfty}. In Appendix \eqref{app:oneloop}
we computed $\langle D_2 \rangle$ and we sketched how to compute
$\langle PD_2 \rangle$. We have not computed the corrected
partition function here explicitly since we do not need it for the
purposes of this paper. However, below we compute it explicitly in
the near-critical regions giving the $SU(2)$ and $SU(2|3)$
sectors.

We use now the result \eqref{final} to compute the one-loop
correction to the Hagedorn temperature. From Eq.~\eqref{Zsingu} we
have that the zeroth order contribution to the partition function
goes like $(T_H-T)^{-1}$ near the Hagedorn temperature, for fixed
chemical potentials. This behavior resists also for the corrected
partition function where now the value of the Hagedorn temperature
is shifted by the higher loop corrections. One can then compute
the one-loop corrected Hagedorn temperature by considering the
pole of ${Z}_{\rm ST}^{(1)}$ in \eqref{final} at the zeroth order
Hagedorn temperature $T^{(0)}_H$. As in the case of zero chemical
potentials~\cite{Spradlin:2004pp} the term proportional to
$\langle PD_2 \rangle$ does not give rise to divergences. Hence,
we get the following formula for the one-loop correction to the
Hagedorn temperature
\begin{equation}
\delta T_H = \lambda \left. \frac{ \langle D_2 \rangle }{ T
\frac{\partial z}{\partial T}} \right|_{T=T_H^{(0)}} \label{1loop}
\end{equation}
for given chemical potentials $\Omega_i$.

Using Eq.~\eqref{1loop}, we compute now the one-loop corrected
Hagedorn temperature for small values of the chemical potentials
$\Omega_i$. To this end, we use the results on $\langle D_2
\rangle$ of Appendix \ref{app:oneloop} to find the following
expression for $\langle D_2 \rangle$ evaluated at the Hagedorn
temperature for small chemical potentials
\begin{eqnarray}
\langle D_2\rangle & = &
\frac{3}{4}\left[1-\frac{\beta_0^2}{18}\sum_{i=1}^3\Omega_i^2-\frac{\beta_0^3}{864}
\left(72-56\sqrt{3}+3\beta_0(41-26\sqrt{3})\right)\sum_{i<j}\Omega_i^2\Omega_j^2
\right. \cr &&\left.-\frac{\beta_0^3}{1296}
\left(72-56\sqrt{3}+3\beta_0(45-26\sqrt{3})\right)\sum_{i=1}^3\Omega_i^4+\CO(\Omega_i^6)\right]
\label{d2small}
\end{eqnarray}
To compute this we used the zeroth order Hagedorn temperature for
small chemical potentials given in Eq.~\eqref{Thag}. Inserting Eq.
\eqref{d2small} in Eq.~\eqref{1loop}, we find that the one-loop
corrected Hagedorn temperature for small chemical potentials is
\begin{equation}
\label{sm1loop}
\begin{array}{c} \ds T_H (\Omega_i) = p_0 + p_1
\sum_{i=1}^3 \Omega_i^2 + p_2 \sum_{i<j} \Omega_i^2 \Omega_j^2 +
p_3 \sum_{i=1}^3 \Omega_i^4  + \CO
(\Omega_i^6 ) \\[5mm] \ds
p_0 = \frac{1}{\beta_0}\left(1+\frac{{\lambda}}{2}\right) +
\CO(\lambda^2) \spa p_1=
-\frac{1}{6\sqrt{3}}\left(1-\frac{{\lambda}}{2}(11-\beta_0\sqrt
3)\right)+ \CO(\lambda^2) \\[4mm] \ds p_2=\frac{\beta_0}{1296}\left[18-5\sqrt{3}+{\lambda}\left(60-\beta_0
\left(72-35\sqrt{3}-\beta_0(69\sqrt{3}-113)\right)\right)\right]+
\CO(\lambda^2) \\[4mm] \ds
p_3=\frac{\beta_0}{2592}\left[18-11\sqrt{3}+{\lambda}\left(60-\beta_0
\left(72-47\sqrt{3}-\beta_0(69\sqrt{3}-122)\right)\right)\right]+
\CO(\lambda^2)
\end{array}
\end{equation}
Note that for zero chemical potentials in Eqs.~\eqref{d2small} and
\eqref{sm1loop} we recover the result of~\cite{Spradlin:2004pp}.

\subsubsection*{The $SU(2)$ sector}

We consider now the near-critical region \eqref{thereg} with
$(\Omega_1,\Omega_2,\Omega_3)=(\Omega,\Omega,0)$. From Section
\ref{sec:QMsec} we know that the single-trace sector of the planar
limit of $U(N)$ $\CN=4$ SYM on $\R \times S^3$ reduces to the
$SU(2)$ sector with single-traces of the form \eqref{singsu2}.
From Section \ref{sec:QMsec} we have furthermore that we can
consider $\tilde{T}=T/(1-\Omega)$ as the effective temperature and
that the one-loop corrected Hamiltonian becomes $H = D_0 +
\tilde{\lambda} D_2$ with $\tilde{\lambda}=\lambda/(1-\Omega)$. In
the following we employ these results to find the corrected
partition function and Hagedorn temperature for this near-critical
region. Note that we assume in the following that $\tilde{\lambda}
\ll 1$.

From Section \ref{sec:freethermal} we have that the zeroth order
contribution to the partition function for the $SU(2)$ sector is
\begin{equation}
Z^{(0)}_{\rm ST}(\tilde{x}) = - \sum_{k=1}^\infty
\frac{\varphi(k)}{k} \log ( 1 - 2\tilde{x}^k )
\end{equation}
We now consider the first correction in $\tilde{\lambda}$ to this
partition function when $\tilde{\lambda} \ll 1$.%
\footnote{The one-loop partition function for the $SU(2)$ sector
is computed previously in~\cite{Spradlin:2004pp}, but we review it
here for completeness, and since we use the same technique below
to compute the first correction for $\tilde{\lambda}\ll 1$ for the
$SU(2|3)$ sector.}
To this end, we use the formula~\cite{Spradlin:2004pp}
\begin{equation}
{Z}_{\rm ST}^{(1)}(\tilde{x})=\log{\tilde x} \sum_{L=1}^{\infty}
\sum_{\tiny \begin{array}{c} k=0 \\[-1mm] (k,L)=1 \end{array}}^{L-1}
\left(\frac{\langle
D_2(\omega^{L+1}\tilde{x}^{L})\rangle}{1-z(\omega^{L+1}\tilde{x}^{L})}+
\delta_{L\neq 1}\langle
PD_2(\omega^{L-k+1}\tilde{x}^{L-k},\omega^{k+1}\tilde{x}^{k})\rangle
\right)\label{1loopsu2}
\end{equation}
In the $SU(2)$ sector the expectation values of $D_2$ and $PD_2$ are
given by~\cite{Spradlin:2004pp}
\be \langle D_2(\tilde{x})\rangle=\tilde{x}^2,~~~~~\langle
PD_2(\tilde{x}_1,\tilde{x}_2)\rangle=-\tilde{x}_1\tilde{x}_2
\label{vevpd2t} \ee
Substituting now those expressions into the formula
\eqref{1loopsu2}, we recover the known result for the one-loop
partition function in the $SU(2)$ sector~\cite{Spradlin:2004pp}
\begin{equation}
\label{Z1su2} Z^{(1)}_{\rm ST}(\tilde{x}) =\log{\tilde
x}\left[\tilde x-\sum_{n=1}^{\infty} \varphi
(n)\tilde{x}^n\left(\frac{1-3\tilde{x}^n}{1-2\tilde{x}^n}\right)\right]
\end{equation}
Similarly to Eq.~\eqref{1loop}, we have that the correction to the
Hagedorn temperature is
\begin{equation}
\label{eqhag2} \delta \tilde{T}_H = \tilde{\lambda} \left.
\frac{\langle D_2 \rangle}{ \tilde{T} \frac{\partial z
(\tilde{x})}{\partial \tilde{T}} }
\right|_{\tilde{T}=\tilde{T}^{(0)}_H}
\end{equation}
where $\tilde{T}^{(0)}_H=1/\log 2$. We used here that $\langle
PD_2 \rangle$ is not divergent, as one can see from
\eqref{vevpd2t}. From Eqs.~\eqref{vevpd2t} and \eqref{eqhag2} we
get then that the corrected Hagedorn temperature for
$\tilde{\lambda}\ll 1$ is
\begin{equation}
\tilde{T}_H =\frac{1}{\log{2}}\left(1+\frac{1}{4}\tilde{\lambda} +
\CO(\tilde{\lambda}^2) \right) \label{hag2}
\end{equation}

It is important to notice that starting instead from the general
expressions for $\langle D_2({x},y_i)\rangle$ and $\langle
PD_2({x},y_i)\rangle$ for $\mathcal{N}=4$ SYM given in Appendix
\ref{app:oneloop} and taking the limit \eqref{declim} precisely
gives the result \eqref{vevpd2t}.%
\footnote{For $\langle D_2 \rangle$ we have from \eqref{expD2} and
\eqref{V0}-\eqref{Vj} in Appendix \ref{app:oneloop} that $V_0$
does not contribute and $V_{j\geq 2} \rightarrow 0$ in the limit
\eqref{declim}, while $V_1 = x^2 y^2$ since only the $x^2
F^{(0,0)}_{[1,0,1]}$ term contributes. For $\langle PD_2 \rangle$
one can take the limit on Eq.~\eqref{pd2} and see that it reduces
to the correct answer.} From this fact one can in turn see that
both the one-loop corrected partition function and Hagedorn
temperature reduces to \eqref{Z1su2} and \eqref{hag2} found above.
This is in accordance with our derivation of the interacting
Hamiltonian in Section \ref{sec:QMsec}.

Finally we note that the two loop corrected Hagedorn temperature
in the $SU(2)$ sector has been considered
in~\cite{Gomez-Reino:2005bq}. However, their result is not
directly applicable in our case, since the two Hamiltonians for
the corrections are different.

\subsubsection*{The $SU(2|3)$ sector}

In the $SU(2|3)$ sector the story is very similar to the one for
the $SU(2)$ sector. We are considering the near-critical region
\eqref{thereg} with
$(\Omega_1,\Omega_2,\Omega_3)=(\Omega,\Omega,\Omega)$. From
Section \ref{sec:QMsec} we know that the single-trace sector of
the planar limit of $U(N)$ $\CN=4$ SYM on $\R \times S^3$ reduces
to the $SU(2|3)$ sector with single-traces of the form
\eqref{singsu23}. The zeroth order single-trace partition function
is
\begin{equation}
Z^{(0)}_{\rm ST}(\tilde{\beta}) = - \sum_{k=1}^\infty
\frac{\varphi(k)}{k} \log \left( 1 - 3\tilde{x}^k - 2(-1)^{k+1}
\tilde{x}^{3k/2} \right)
\end{equation}
We compute the first correction in $\tilde{\lambda}$ to this
partition function when $\tilde{\lambda} \ll 1$ using again
Eq.~\eqref{1loopsu2}.  Using that the dilatation operator is given
by \eqref{D2su23} we find
\begin{equation}
\begin{array}{rcl}
\langle D_2(\tilde{x})\rangle&=& \ds
3\tilde{x}^2+6\tilde{x}^{5/2}+3\tilde{x}^3 \\[2mm] \langle
PD_2(\tilde{x}_1,\tilde{x}_2)\rangle &=& \ds
-3\tilde{x}_1\tilde{x}_2+3\tilde{x}_1^{3/2}\tilde{x}_2^{3/2}
-3\left[\tilde{x}_1^{3/2}\tilde{x}_2 +\tilde{x}_1\tilde{x}_2^{3/2}
\right]
\end{array}
\label{vevpd23t}
\end{equation}
As for the $SU(2)$ sector, these results can be recovered using
the expressions for $\langle D_2({x},y_i)\rangle$ and $\langle
PD_2({x},y_i)\rangle$ for $\mathcal{N}=4$ SYM given in Appendix
\ref{app:oneloop} and taking the limit \eqref{declim}. Inserting
the previous expressions in Eq.~\eqref{1loopsu2} we get that the
one-loop partition function in the $SU(2|3)$ sector is given by
\begin{equation}
\begin{array}{rcl} \ds
{Z}^{(1)}_{\rm{ST}}(\tilde{x}) &=& \ds \log{\tilde x}  \left[
3\tilde{x} + 3\tilde{x}^{3/2} - 3 \sum_{L=2}^\infty  \right.
\sum_{\tiny
\begin{array}{c} k=0 \\[-1mm] (k,L)=1 \end{array}}^{L-1}
\left( (-1)^{L-k+1} \tilde{x}^{(3L-k)/2} + (-1)^{k+1}
\tilde{x}^{L+k/2} \right)
\\[7mm] && \ \ \ \ \ \ \ \ds \left.
 - 3 \sum_{n=1}^{\infty} \varphi (n) \tilde{x}^n
\frac{ 1 - (-1)^n \tilde{x}^{n/2} - 4 \tilde{x}^n + 7 (-1)^n
\tilde{x}^{3n/2} - 3
\tilde{x}^{2n}}{1-3\tilde{x}^n-2(-1)^{n+1}\tilde{x}^{3n/2}}
\right]
\end{array}
\end{equation}
Using now Eqs.~\eqref{eqhag2} and \eqref{vevpd23t} we get for the
one-loop corrected Hagedorn temperature the following result
\be \tilde{T}_H=\frac{1}{\log{4}}\left(1+ \frac{3}{8}
\tilde{\lambda}+ \CO (\tilde{\lambda}^2) \right) \label{hagsu23}
\ee
One can check that only the $\langle D_2(\tilde{x})\rangle$ part
of the one-loop partition function contributes to this.

\section{Discussion and conclusions}
\label{sec:concl}

In this paper we have found that thermal $\CN=4$ SYM on $\R \times
S^3$ greatly reduces near the critical points
$(T,\Omega_1,\Omega_2,\Omega_3)=(0,1,0,0)$,
$(T,\Omega_1,\Omega_2,\Omega_3)=(0,1,1,0)$ and
$(T,\Omega_1,\Omega_2,\Omega_3)=(0,1,1,1)$. We identified the
three quantum mechanical theories that $\CN=4$ SYM reduces to, and
in particular we showed that the Hilbert spaces correspond to a
half-BPS sector and the $SU(2)$ and $SU(2|3)$ sectors of $\CN=4$
SYM. We found the Hamiltonian for these three theories and we saw
that the one-loop correction to the dilatation operator has a
special significance in this. The existence of these quantum
mechanical sectors of $\CN=4$ SYM could prove highly useful.
Through the AdS/CFT correspondence the thermodynamics of $\CN=4$
SYM is linked to the Hagedorn transition in string theory, and
since for instance the $SU(2)$ sector is greatly reduced in
complexity compared to the full $\CN=4$ SYM, we can get a much
better handle on the behavior of $\CN=4$ SYM in this particular
near-critical region than on $\CN=4$ SYM with zero chemical
potentials.

For $N=\infty$ we found that the near-critical regions giving the
$SU(2)$ and $SU(2|3)$ sectors can be described in terms of spin
chain theories. In particular the $SU(2)$ sector corresponds to a
ferromagnetic $XXX_{1/2}$ Heisenberg spin chain to leading order
(or exactly, if we take the limit of Section \ref{sec:declim}).
This provides a very different realization of spin chains for the
planar limit of $\CN=4$ SYM on $\R \times S^3$ than in the study
of integrability
\cite{Minahan:2002ve,Beisert:2003tq,Beisert:2003yb}. In terms of
integrability, the $SU(2)$ and $SU(2|3)$ sectors are closed
subsectors of the conjectured complete $\CN=4$ spin chain, i.e.
they decouple to all orders in perturbation
theory~\cite{Beisert:2003ys}. However, it is not clear that this
decoupling holds at strong coupling
\cite{Callan:2003xr,Minahan:2005jq}. Instead, in the limit of this
paper we have an effective reduction of $\CN=4$ SYM to the $SU(2)$
and $SU(2|3)$ sectors which does not rely on the $SU(2)$ and
$SU(2|3)$ sectors being closed in the sense of having interactions
with the other operators of $\CN=4$ SYM. Any such interaction
would in any case be suppressed in the near-critical regions that
we consider. It would therefore be interesting to consider if our
decoupling of the $SU(2)$ and $SU(2|3)$ sectors corresponds to a
similar decoupling for thermal string theory on $\ads_5 \times
S^5$ with near-critical chemical potentials.

It is intriguing to compare our limit to the pp-wave limits of
$\ads_5 \times S^5$ \cite{Berenstein:2002jq}. It is not hard to
see that the near-critical region giving us the reduction to the
$SU(2)$ sector has some similarities with the pp-wave limit of
\cite{Bertolini:2002nr} since we keep only states with
$D_0=R_1+R_2$. It is clear that to connect to the limit of $\CN=4$
SYM found in \cite{Bertolini:2002nr} we need to consider only a
subsector of the pp-wave string theory of \cite{Bertolini:2002nr}.
This seems possible to achieve by turning on the appropriate
chemical potential. This would be interesting to study since we
have a Hagedorn transition both in the gauge theory side and on
the pp-wave side \cite{PandoZayas:2002hh}.

Another interesting direction to pursue would be to compare our
results on the Hagedorn temperature as a function of the chemical
potential to the Hawking-Page transition
\cite{Hawking:1982dh,Witten:1998zw} with chemical potentials
\cite{Buchel:2003re}. With the chemical potentials set to zero we
have a consistent picture that the Hagedorn transition is a first
order transition both for weak coupling $\lambda \ll 1$
\cite{Aharony:2003sx,Aharony:2005bq} and for strong coupling
$\lambda \gg 1$ \cite{Barbon:1998ix} where it is mapped to the
Hawking-Page transition. It would be interesting to see whether
the picture is equally consistent once the chemical potentials are
turned on.

Finally, we note that we expect similar decoupled quantum
mechanical sectors in other supersymmetric gauge theories with
R-symmetry, in regions with near-critical chemical potentials.

\section*{Acknowledgments}

We thank P. Di Vecchia, G. Grignani, C. Kristjansen and N. Obers
for useful discussions and H. Osborn for useful correspondence. We
thank KITP for hospitality while part of this work was completed.
This research was supported in part by the National Science
Foundation under Grant No. PHY99-0794. The work of M.O. is
supported in part by the European Community's Human Potential
Programme under contract MRTN-CT-2004-005104 `Constituents,
fundamental forces and symmetries of the universe'.

\begin{appendix}

\section{Oscillator representation of $\CN=4$ SYM}
\label{app:osc}

In the oscillator representation of $\CN=4$ SYM
\cite{Gunaydin:1984fk,Beisert:2003jj} we can write all the
gauge-invariant operators using two bosonic oscillators
$\mathbf{a}^{\alpha}$, $\mathbf{b}^{\dot\alpha}$,
$\alpha,\dot\alpha=1,2$, and one fermionic oscillator
$\mathbf{c}^{a}$, $a=1,2,3,4$, with the commutation relations
\begin{equation}
\left[\mathbf{a}^{\alpha},{\mathbf{a}}^{\dagger}_{\beta}\right]=\delta^{\alpha}_{\beta}~~~~
\left[\mathbf{b}^{\dot\alpha},{\mathbf{b}}^{\dagger}_{\dot\beta}\right]=\delta^{\dot\alpha}_{\dot\beta}~~~~
\left\{\mathbf{c}^{a},{\mathbf{c}}^{\dagger}_{b}\right\}=\delta^a_b
\end{equation}
In terms of these oscillators, the set of letters $\CA$ of $\CN=4$
SYM is given by
\begin{equation}
\begin{array}{c}
\phi \ :\ ({\mathbf{c}}^{\dagger})^2 |0 \rangle
 \mbox{ repr. } [0,1,0]_{(0,0)} \\[1mm]
\psi \ :\ {\mathbf{a}}^{\dagger} {\mathbf{c}}^{\dagger} |0 \rangle
\mbox{ repr. } [0,0,1]_{(\frac{1}{2},0)} \spa \bar{\psi} \ :\
{\mathbf{b}}^{\dagger} ({\mathbf{c}}^{\dagger})^3 |0 \rangle
 \mbox{ repr. }  [1,0,0]_{(0,\frac{1}{2})}  \\[1mm]
F \ :\  ({\mathbf{a}}^{\dagger})^2 |0 \rangle \mbox{ repr. }
[0,0,0]_{(1,0)}  \spa \bar{F} \ :\ ({\mathbf{b}}^{\dagger})^2 |0
\rangle \mbox{ repr. }
[0,0,0]_{(0,1)}  \\[1mm]
D  \ :\ {\mathbf{a}}^{\dagger} {\mathbf{b}}^{\dagger} \mbox{ repr.
} [0,0,0]_{(\frac{1}{2},\frac{1}{2})}
\end{array}
\end{equation}
where $F$ is the field strength, $\psi$ the fermions and $\phi$
the scalars. Moreover $D$ is the covariant derivative. One can
then generate $\CA$ by acting with $D^k$. Note that we also
specified the representation under $SU(4)\times SO(4)$ that the
fields are in, for example $[0,0,1]_{(1/2,0)}$ corresponds to the
$[0,0,1]$ of $SU(4)$ and the $(\frac{1}{2},0)$ of $SO(4)$.

Write now the number operators as $a^\alpha =
{\mathbf{a}}^{\dagger}_{\alpha} \mathbf{a}^{\alpha}$,
$b^{\dot\alpha} = {\mathbf{b}}^{\dagger}_{\dot\alpha}
\mathbf{b}^{\dot\alpha}$ and $c^a = {\mathbf{c}}^{\dagger}_{a}
\mathbf{c}^{a}$, where it should be understood that there are no
sums over the indices. We define then the operators
\begin{equation}
\begin{array}{rcl}
\ds C &=& \ds 1 - \frac{1}{2} (a^1+a^2) + \frac{1}{2} (b^1+b^2) -
\frac{1}{2} (c^1+c^2+c^3+c^4) \\[3mm]
\ds D_0 &=& \ds 1 + \frac{1}{2} (a^1+a^2 + b^1+b^2)
\end{array}
\end{equation}
Here $C$ is the central charge which should be annihilated on
physical states, while $D_0$ is the dilatation operator in free
$\CN=4$ SYM. The three R-charges are
\begin{equation}
R_1 = \frac{1}{2} (c^1-c^2-c^3+c^4) \spa R_2 = \frac{1}{2}
(-c^1+c^2-c^3+c^4) \spa R_3 = \frac{1}{2} (-c^1-c^2+c^3+c^4)
\end{equation}
We can now write the letter partition function as
\begin{equation}
\label{osclettpart}
\begin{array}{l}
\ds z(x,y_1,y_2,y_3) = {\tr}_\CA \left( x^{D_0} y_1^{R_1}
y_2^{R_2} y_3^{R_3} \right) = \sum_{a_1,a_2,b_1,b_2=0}^\infty
\sum_{c^1,c^2,c^3,c^4=0}^1 \delta (C) x^{D_0} y_1^{R_1} y_2^{R_2}
y_3^{R_3} \\[5mm] \ds = \sum_{a,b=0}^\infty (a+1)(b+1)
\sum_{c^1,c^2,c^3,c^4=0}^1 \delta \Big(2-a+b-\sum_{a=1}^4 c^a
\Big) x^{1+\frac{1}{2}(a+b)} y_1^{R_1} y_2^{R_2} y_3^{R_3}
\end{array}
\end{equation}
It is straightforward to see that this gives the letter partition
function \eqref{lettpart} computed in Section \ref{sec:partfct}.
Note that we defined $a=a^1+a^2$ and $b=b^1+b^2$ in
\eqref{osclettpart}.

We consider now the decoupling limits of Section \ref{sec:nearcrit}.
Consider first the case in which $\Omega_1=\Omega_2=\Omega$,
$\Omega_3=0$ and hence $R=R_1+R_2$. Taking the limit \eqref{xylim},
i.e. with $\tilde{x} \equiv xy$ fixed and $x\rightarrow 0$,
$y=\exp(\beta \Omega)$, it is easy to see that only the sector with
$D_0=R$ survives. Using the above formulas we see that since
$R=R_1+R_2=-c^3+c^4$ the limit \eqref{xylim} corresponds to
inserting the kronecker delta $\delta (2+a+b+2c^3-2c^4)$ into the
sum in \eqref{osclettpart}. This kronecker delta-function can
clearly only be $1$ provided $a=b=c^3=0$ and $c^4=1$, since all the
number operators are positive and the fermionic number operators
only take the values $0$ and $1$. We are thus in the sector given by
\begin{equation}
a^1=a^2=b^1=b^2=c^3=0 \spa c^4=1
\end{equation}
and it is easy to see that the only states in this sector are
${\mathbf{c}}^{\dagger}_1 {\mathbf{c}}^{\dagger}_4 |0 \rangle$ and
${\mathbf{c}}^{\dagger}_2 {\mathbf{c}}^{\dagger}_4 |0 \rangle$,
corresponding to the two complex scalars $Z$ and $X$. This is
clearly the $SU(2)$ sector, as defined in \cite{Beisert:2003jj},
and the partition function is indeed easily found from
\eqref{osclettpart} to reduce to \eqref{zcII}.

Consider instead the case in which
$\Omega_1=\Omega_2=\Omega_3=\Omega$ and hence $R=R_1+R_2+R_3$.
Taking the limit \eqref{xylim} we see again that only the sector
with $D_0=R$ remains. Using that
$R=R_1+R_2+R_3=\frac{1}{2}(-c^1-c^2-c^3+3c^4)$ we see that this
limit corresponds to inserting $\delta(2+a+b+c^1+c^2+c^3-3c^4)$
into the sum in \eqref{osclettpart}. It is clear that this
kronecker delta only can be non-zero provided $c^4=1$. If we
consider the case $b=1$ we see that then we need
$a=c^1=c^2=c^3=0$, but that is not a physical state. This means
that $b=0$ and that $a+c^1+c^2+c^3=1$, which is equivalent to
stating that $b=0$ and $C=0$. We are thus in the sector given by
\begin{equation}
b^1=b^2=0 \spa c^4 = 1
\end{equation}
The physical states in this sector are ${\mathbf{c}}^{\dagger}_1
{\mathbf{c}}^{\dagger}_4 |0 \rangle$, ${\mathbf{c}}^{\dagger}_2
{\mathbf{c}}^{\dagger}_4 |0 \rangle$ and ${\mathbf{c}}^{\dagger}_3
{\mathbf{c}}^{\dagger}_4 |0 \rangle$, corresponding to the three
complex scalars $Z$, $X$ and $W$, and ${\mathbf{a}}^{\dagger}_1
{\mathbf{c}}^{\dagger}_4 |0 \rangle$ and ${\mathbf{a}}^{\dagger}_2
{\mathbf{c}}^{\dagger}_4 |0 \rangle$ corresponding to the two
complex fermions $\chi_1$ and $\chi_2$. This is clearly the
$SU(2|3)$ sector defined in \cite{Beisert:2003jj}. Furthermore, it
is straightforward to find that the partition function
\eqref{osclettpart} reduces to \eqref{zcIII}.

\section{The $XXX_{1/2}$ Heisenberg spin chain}
\label{app:XXX}

For convenience we briefly review here some essential facts of the
$XXX_{1/2}$ Heisenberg spin chain. We are considering a periodic
spin chain of length $L$, so that the Hilbert space of the spin
chain is spanned by states with $M$ down-spins and $L-M$ up-spins,
$0\leq M \leq L$. Thus, the Hilbert space has dimension $2^L$. The
Hamiltonian of a one-dimensional $XXX_{1/2}$ Heisenberg spin chain
is traditionally defined as
\begin{equation}
\label{HXXX} H = J \sum_{i=1}^L \left( \vec{S}_i \cdot
\vec{S}_{i+1} - \frac{1}{4} \right)
\end{equation}
where $\vec{S}_i$ acts on the $i$'th spin as $\vec{\sigma}/2$,
i.e. with $\vec{\sigma}$ being the Pauli matrices.  To find the
eigenvalues and eigenstates of the Hamiltonian one uses the Bethe
ansatz \cite{Bethe:1931hc} (see for example \cite{Berruto:1999ga}
for specific examples of spectra for $L=4,6$). In
\cite{Faddeev:1981ip} the full spectrum has been found in the
thermodynamic limit $L \rightarrow \infty$. Defining the total
spin
\begin{equation}
\vec{S} = \sum_{i=1}^L \vec{S}_i
\end{equation}
we have that $[H,\vec{S}]=\vec{0}$. This means that any eigenstate
of $H$ is part of a spin multiplet with respect to $\vec{S}$.

If we have $J < 0$ the Hamiltonian \eqref{HXXX} is describing a
ferromagnet. The ferromagnetic vacua are the states with
eigenvalue zero of $H$. These are totally symmetrized states with
$M$ down-spins and $L-M$ up-spins, $0\leq M \leq L$. Clearly there
are $L+1$ such states and they in fact make up a $L+1$ dimensional
representation with respect to $\vec{S}$.

If we have instead that $J > 0$ the Hamiltonian \eqref{HXXX} is
describing an antiferromagnet. The antiferromagnetic vacuum state
is a unique state with $L/2$ up-spins and $L/2$ down-spins
(assuming $L$ even). It is a singlet with respect to $\vec{S}$.

\section{Computations for one-loop partition function}
\label{app:oneloop}

In this appendix we derive the expression for $\langle
D_2(x,y_i)\rangle$ used in Section \ref{sec:oneloop} to compute
the one-loop correction to the Hagedorn temperature. We also
briefly discuss how to compute $\langle P
D_2(x_1,y_{i(1)},x_2,y_{i(2)})\rangle$.

From the definition \eqref{Oy} of $\langle D_2(x,y_i)\rangle$ we
have that it corresponds to the expectation value of $D_2$ acting
on the product of two copies of the singleton representation
$\mathcal{A}\times \mathcal{A}$. To compute $\langle
D_2(x,y_i)\rangle$ we can then employ the fact that it commutes
with the two-letter Casimir of $PSU(2,2|4)$ on $ \mathcal{A}\times
\mathcal{A}$ \cite{Beisert:2003jj}. To this end, we use the
following modules of $PSU(2,2|4)$
\cite{Dolan:2002zh,Bianchi:2003wx}%
\begin{equation}
\mathcal{A}=\mathcal{B}^{\frac{1}{2},\frac{1}{2}}_{[0,1,0]_{(0,0)}}~~~
\CV_0=\mathcal{B}^{\frac{1}{2},\frac{1}{2}}_{[0,2,0]_{(0,0)}}~~~
\CV_1=\mathcal{B}^{\frac{1}{4},\frac{1}{4}}_{[1,0,1]_{(0,0)}}~~~
\CV_j=\mathcal{C}^{1,1}_{[0,0,0]_{(\frac{j}{2}-1,\frac{j}{2}-1)}}~\mbox{for
$j\geq 2$.} \label{modules}
\end{equation}
Here we wrote the modules in the notation of \cite{Dolan:2002zh}.
For each module it is written what superconformal primary operator
the representation is generated from, e.g. for $\CV_1$ it is
$[1,0,1]_{(0,0)}$ which is the primary operator in the $[1,0,1]$
representation of $SU(4)$ and in the singlet $(0,0)$ of
$SU(2)\times SU(2)$. We have then that $\CA \times \CA =
\sum_{j=0}^\infty \CV_j$ and that the eigenvalue of $D_2$ in
$\CV_j$ is given by the harmonic number
$h(j)=\sum_{n=1}^j\frac{1}{n}$~\cite{Beisert:2003jj}. We can
therefore compute $\langle D_2(x,y_i)\rangle$ by computing
${\tr}_{\CV_j} ( x^{D_0} \prod_{i=1}^3 y_i^{R_i} )$. This can be
done using the tables for the modules (\ref{modules}) presented
in~\cite{Dolan:2002zh,Bianchi:2003wx,Spradlin:2004pp}. We define
\begin{equation}
F_{[k,p,q]}^{(j_1,j_2)} = (2j_1+1)(2j_2+1) W_{[k,p,q]} \spa
W_{[k,p,q]} \equiv {\tr}_{[k,p,q]} ( y_i^{R_i} )
\end{equation}
We see that $W_{[k,p,q]}$ is the weighted sum of the weights of
$[k,p,q]$. For the specific representations we have
\begin{equation}
\label{v}
\begin{array}{c} \ds
W_{[0,0,0]}=1,~~~~W_{[0,1,0]}=\sum_{i=1}^3\big(y_i+y_i^{-1}\big),~~~~
W_{\left([1,0,0]+[0,0,1]\right)}=
\prod_{i=1}^3\big(y_i^{1/2}+y_i^{-1/2}\big)
\\[2mm] \ds
W_{\left([1,1,0]+[0,1,1]\right)}=
\prod_{i=1}^3\big(y_i^{1/2}+y_i^{-1/2}\big)
\left[\sum_{j=1}^3\big(y_j+y_j^{-1}\big) -1\right]
\\[5mm] \ds
W_{[0,2,0]}=\sum_{1\leq i\leq j \leq 3}
\big(y_i+y_i^{-1}\big)\big(y_j+y_j^{-1}\big)-4,~~~
W_{[1,0,1]}=\sum_{1 \leq i< j \leq
3}\big(y_i+y_i^{-1}\big)\big(y_j+y_j^{-1}\big) + 3
\\[2mm] \ds
W_{\left([2,0,0]+[0,0,2]\right)}= 2 \sum_{i=1}^3\big(
y_i+y_i^{-1}\big)+\prod_{i=1}^3\big(y_i+y_i^{-1}\big)
\end{array}
\end{equation}
From the above we can now compute $V_j(x,y_i) \equiv (1-x)^4
{\tr}_{\CV_j} ( x^{D_0} \prod_{i=1}^3 y_i^{R_i} )$. We get
\bea && V_0 = x^2F_{[0,2,0]}^{(0,0)} +x^{\frac{5}{2}} \Big[
F_{[1,1,0]}^{(0,\frac{1}{2})} + F_{[0,1,1]}^{(\frac{1}{2},0)}
\Big] + x^3 \Big[ F_{[2,0,0]}^{(0,0)} + F_{[0,0,2]}^{(0,0)} +
F_{[0,1,0]}^{(1,0)} + F_{[0,1,0]}^{(0,1)}  +
F_{[1,0,1]}^{(\frac{1}{2},\frac{1}{2})} \Big] \cr &&
 + x^{\frac{7}{2}}
 \Big[ F_{[0,0,1]}^{(\frac{1}{2},0)}+F_{[1,0,0]}^{(0,\frac{1}{2})}+
F_{[1,0,0]}^{(1,\frac{1}{2})}+F_{[0,0,1]}^{(\frac{1}{2},1)} \Big]
+  x^4 \Big[ 2F_{[0,0,0]}^{(0,0)} + F_{[0,0,0]}^{(1,1)}
-F_{[1,0,1]}^{(0,0)} \Big] \cr &&  - x^{\frac{9}{2}} \Big[
F_{[1,0,0]}^{(\frac{1}{2},0)} + F_{[0,0,1]}^{(0,\frac{1}{2})}
\Big] - x^5 F_{[0,0,0]}^{(\frac{1}{2},\frac{1}{2})} \label{V0}
\eea
\bea && V_1 = x^2F_{[1,0,1]}^{(0,0)} +x^{\frac{5}{2}}\Big[
F_{[1,0,0]}^{(\frac{1}{2},0)}+F_{[0,0,1]}^{(0,\frac{1}{2})}
+F_{[1,1,0]}^{(0,\frac{1}{2})}+F_{[0,1,1]}^{(\frac{1}{2},0)} \Big]
 \cr && + x^3 \Big[ 2F_{[0,1,0]}^{(0,0)} +
F_{[0,1,0]}^{(1,0)} +F_{[0,1,0]}^{(0,1)} +F_{[0,0,2]}^{(1,0)}
+F_{[2,0,0]}^{(0,1)} +F_{[0,0,0]}^{(\frac{1}{2},\frac{1}{2})} +
 F_{[1,0,1]}^{(\frac{1}{2},\frac{1}{2})}+
 F_{[0,2,0]}^{(\frac{1}{2},\frac{1}{2})} \Big]
\cr && + x^{\frac{7}{2}}\Big[ F_{[0,0,1]}^{(\frac{1}{2},0)} +
F_{[1,0,0]}^{(0,\frac{1}{2})} +F_{[0,0,1]}^{(\frac{3}{2},0)}  +
 F_{[1,0,0]}^{(0,\frac{3}{2})}
+F_{[1,0,0]}^{(1,\frac{1}{2})}+F_{[0,0,1]}^{(\frac{1}{2},1)} +
F_{[0,1,1]}^{(1,\frac{1}{2})}
 +F_{[1,1,0]}^{(\frac{1}{2},1)}
\Big]\cr && + x^4\Big[ F_{[0,0,0]}^{(1,0)}+F_{[0,0,0]}^{(0,1)}
+F_{[0,1,0]}^{(\frac{1}{2},\frac{3}{2})}
+F_{[0,1,0]}^{(\frac{3}{2},\frac{1}{2})} + F_{[0,0,0]}^{(1,1)} +
F_{[1,0,1]}^{(1,1)}-F_{[0,0,0]}^{(0,0)}-F_{[1,0,1]}^{(0,0)}-F_{[0,2,0]}^{(0,0)}
\Big] \cr && + x^{\frac{9}{2}}\Big[
F_{[1,0,0]}^{(\frac{3}{2},1)}+F_{[0,0,1]}^{(1,\frac{3}{2})} -
F_{[0,0,1]}^{(0,\frac{1}{2})}
-F_{[1,0,0]}^{(\frac{1}{2},0)}-F_{[1,1,0]}^{(0,\frac{1}{2})}
-F_{[0,1,1]}^{(\frac{1}{2},0)}
\Big] \cr && + x^5\Big[
F_{[0,0,0]}^{(\frac{3}{2},\frac{3}{2})}-F_{[0,1,0]}^{(0,1)} -
F_{[0,1,0]}^{(1,0)} -
F_{[0,0,0]}^{(\frac{1}{2},\frac{1}{2})}-F_{[1,0,1]}^{(\frac{1}{2},\frac{1}{2})}
\Big]- x^{\frac{11}{2}}\Big[
F_{[0,0,1]}^{(\frac{1}{2},1)}+F_{[1,0,0]}^{(1,\frac{1}{2})} \Big]
-x^6 F_{[0,0,0]}^{(1,1)} \cr &&\label{V1} \eea
\bea && V_j  = x^j F_{[0,0,0]}^{(\frac{j}{2}-1,\frac{j}{2}-1)}
+x^{j+\frac{1}{2}}\Big[ F_{[1,0,0]}^{(\frac{j}{2},\frac{j-1}{2})}
+F_{[0,0,1]}^{(\frac{j-1}{2},\frac{j}{2})} \Big] + x^{j+1}\Big[
F_{[0,1,0]}^{(\frac{j}{2},\frac{j}{2}-1)}
+F_{[0,1,0]}^{(\frac{j}{2}-1,\frac{j}{2})}
 +F_{[1,0,1]}^{(\frac{j-1}{2},\frac{j-1}{2})}
\cr &&
+F_{[0,0,0]}^{(\frac{j-1}{2},\frac{j-1}{2})}
-F_{[0,0,0]}^{(\frac{j-3}{2},\frac{j-3}{2})} \Big] +
x^{j+\frac{3}{2}}\Big[ F_{[0,0,1]}^{(\frac{j+1}{2},\frac{j}{2}-1)}
+F_{[1,0,0]}^{(\frac{j}{2}-1,\frac{j+1}{2})}
+F_{[0,0,1]}^{(\frac{j-1}{2},\frac{j}{2})}
+F_{[1,0,0]}^{(\frac{j}{2},\frac{j-1}{2})}
\cr &&
+F_{[0,1,1]}^{(\frac{j}{2},\frac{j-1}{2})} +
F_{[1,1,0]}^{(\frac{j-1}{2},\frac{j}{2})}-F_{[1,0,0]}^{(\frac{j}{2}-1,\frac{j-3}{2})}
-F_{[0,0,1]}^{(\frac{j-3}{2},\frac{j}{2}-1)} \Big]+ x^{j+2}\Big[
F_{[0,0,0]}^{(\frac{j}{2}+1,\frac{j}{2}-1)}
+F_{[0,0,0]}^{(\frac{j}{2}-1,\frac{j}{2}+1)}
\cr &&
+F_{[0,1,0]}^{(\frac{j+1}{2},\frac{j-1}{2})}+F_{[0,1,0]}^{(\frac{j-1}{2},\frac{j+1}{2})}
+ F_{[0,0,0]}^{(\frac{j}{2},\frac{j}{2})} +
F_{[1,0,1]}^{(\frac{j}{2},\frac{j}{2})}+F_{[0,2,0]}^{(\frac{j}{2},\frac{j}{2})}+
F_{[0,0,2]}^{(\frac{j+1}{2},\frac{j-1}{2})}
+F_{[2,0,0]}^{(\frac{j-1}{2},\frac{j+1}{2})}
\cr &&
-F_{[0,0,0]}^{(\frac{j}{2}-1,\frac{j}{2}-1)}-
F_{[1,0,1]}^{(\frac{j}{2}-1,\frac{j}{2}-1)}-F_{[0,1,0]}^{(\frac{j-1}{2},\frac{j-3}{2})}
-F_{[0,1,0]}^{(\frac{j-3}{2},\frac{j-1}{2})} \Big] +
x^{j+\frac{5}{2}}\Big[ F_{[1,0,0]}^{(\frac{j-1}{2},\frac{j}{2}+1)}
+F_{[0,0,1]}^{(\frac{j}{2}+1,\frac{j-1}{2})}
\cr &&
+F_{[0,1,1]}^{(\frac{j+1}{2},\frac{j}{2})}
+F_{[1,1,0]}^{(\frac{j}{2},\frac{j+1}{2})}
+F_{[1,0,0]}^{(\frac{j+1}{2},\frac{j}{2})}
+F_{[0,0,1]}^{(\frac{j}{2},\frac{j+1}{2})}
-F_{[0,0,1]}^{(\frac{j}{2},\frac{j-3}{2})}
-F_{[1,0,0]}^{(\frac{j-3}{2},\frac{j}{2})}
-F_{[1,1,0]}^{(\frac{j}{2}-1,\frac{j-1}{2})}
\cr &&
-F_{[0,1,1]}^{(\frac{j-1}{2},\frac{j}{2}-1)}
-F_{[0,0,1]}^{(\frac{j}{2}-1,\frac{j-1}{2})}
-F_{[1,0,0]}^{(\frac{j-1}{2},\frac{j}{2}-1)} \Big]
+x^{j+3}\Big[F_{[0,1,0]}^{(\frac{j}{2}+1,\frac{j}{2})}
+F_{[0,1,0]}^{(\frac{j}{2},\frac{j}{2}+1)}
+F_{[1,0,1]}^{(\frac{j+1}{2},\frac{j+1}{2})}
\cr &&
+F_{[0,0,0]}^{(\frac{j+1}{2},\frac{j+1}{2})}
-F_{[0,0,0]}^{(\frac{j+1}{2},\frac{j-3}{2})}
-F_{[0,0,0]}^{(\frac{j-3}{2},\frac{j+1}{2})}
-F_{[0,1,0]}^{(\frac{j}{2},\frac{j}{2}-1)}
-F_{[0,1,0]}^{(\frac{j}{2}-1,\frac{j}{2})}
-F_{[0,0,2]}^{(\frac{j}{2},\frac{j}{2}-1)}
-F_{[2,0,0]}^{(\frac{j}{2}-1,\frac{j}{2})}
\cr &&
-F_{[0,2,0]}^{(\frac{j-1}{2},\frac{j-1}{2})}
-F_{[0,0,0]}^{(\frac{j-1}{2},\frac{j-1}{2})}
-F_{[1,0,1]}^{(\frac{j-1}{2},\frac{j-1}{2})} \Big] -
x^{j+\frac{7}{2}}\Big[ F_{[0,0,1]}^{(\frac{j+1}{2},\frac{j}{2}+1)}
+F_{[1,0,0]}^{(\frac{j}{2}+1,\frac{j+1}{2})}
-F_{[0,0,1]}^{(\frac{j+1}{2},\frac{j}{2}-1)}
\cr &&
-F_{[1,0,0]}^{(\frac{j}{2}-1,\frac{j+1}{2})}
-F_{[1,0,0]}^{(\frac{j}{2},\frac{j-1}{2})}
-F_{[0,0,1]}^{(\frac{j-1}{2},\frac{j}{2})}
-F_{[0,1,1]}^{(\frac{j}{2},\frac{j-1}{2})}
-F_{[1,1,0]}^{(\frac{j-1}{2},\frac{j}{2})} \Big] -x^{j+4} \Big[
F_{[0,0,0]}^{(\frac{j}{2}+1,\frac{j}{2}+1)}
-F_{[0,1,0]}^{(\frac{j+1}{2},\frac{j-1}{2})}
\cr &&
-F_{[0,1,0]}^{(\frac{j-1}{2},\frac{j+1}{2})}
-F_{[1,0,1]}^{(\frac{j}{2},\frac{j}{2})}-
F_{[0,0,0]}^{(\frac{j}{2},\frac{j}{2})} \Big]  -
x^{j+\frac{9}{2}}\Big[ F_{[1,0,0]}^{(\frac{j+1}{2},\frac{j}{2})}
+F_{[0,0,1]}^{(\frac{j}{2},\frac{j+1}{2})} \Big]
-x^{j+5}F_{[0,0,0]}^{(\frac{j+1}{2},\frac{j+1}{2})} \label{Vj}
\eea
A nice check of the formulas derived for $V_j(x,y_i)$ is given by
the following equality
\begin{equation}
\sum_{j=0}^{\infty}\frac{V_j(x,y_i)}{(1-x)^4}=z(x,y_i)^2
\label{squared}
\end{equation}
where $z(x,y_i)$ is the letter partition function
\eqref{lettpart}. From the above, we have then
\begin{equation}
\langle D_2(x,y_i)\rangle={\tr}_{\mathcal{A} \times
\mathcal{A}}\left[x^{D_0}\prod_{i=1}^3 y_i^{R_i}D_2 \right] =
\sum_{j=0}^{\infty}h(j)\frac{V_j(x,y_i)}{(1-x)^4} \label{expD2}
\end{equation}
Using Eqs.~\eqref{V0}, \eqref{V1} and \eqref{Vj} for $V_j$ one can
then obtain the expression for $\langle D_2(x,y_i)\rangle$. We do
not write the result here, since it is a highly complicated
expression. Instead we use in Section \ref{sec:oneloop}
Eq.~\eqref{expD2} to find $\langle D_2(x,y_i)\rangle$ for small
chemical potentials and for near-critical chemical potentials.


\subsubsection*{Oscillator representation of $\langle PD_2\rangle$}

We explain here briefly how to compute $\langle
PD_2(x_1,y_{i(1)},x_2,{y}_{i(2)})\rangle$ defined in \eqref{POy}.
We do not compute the resulting expression here due to the fact
that it does not contribute to the correction to the Hagedorn
temperature.

It is not possible to employ the same technique used above for
$\langle D_2 \rangle$ to compute $\langle PD_2 \rangle$, since
$PD_2$, unlike $D_2$, does not commute with the two-letter
$PSU(2,2|4)$ Casimir \cite{Spradlin:2004pp}. Instead, we use the
oscillator representation of $\CN=4$ reviewed in Appendix
\ref{app:osc} to write down an expression for $\langle
PD_2\rangle$. Following \cite{Spradlin:2004pp}, we find
\begin{equation}
\label{pd2} \begin{array}{l} \ds \langle P
D_2(x_1,y_{i(1)},x_2,{y}_{i(2)})\rangle=
\sum_{a^\alpha_{(i)},b^{\dot{\alpha}}_{(i)}=0}^\infty
\sum_{c^a_{(i)}=0}^1 \prod_{i=1}^2 \delta (C_{(i)})
x_{(i)}^{D_{0(i)}} y_{1(i)}^{R_{1(i)}} y_{2(i)}^{R_{2(i)}}
y_{3(i)}^{R_{3(i)}}
\\[7mm] \ds
\times \sum_{k,k',p,p'=0}^{\infty}
 \left( \begin{array}{c} a^1_{(i)} \\ k \end{array} \right)
 \left( \begin{array}{c} a^2_{(i)} \\ k' \end{array} \right)
 \left( \begin{array}{c} b^1_{(i)} \\ p \end{array} \right)
 \left( \begin{array}{c} b^2_{(i)}\\ p' \end{array} \right)
 \sum_{l_1,l_2,l_3,l_4=0}^1
 \prod_{a=1}^4
 \left( \begin{array}{c} c^a_{(i)} \\ l_a \end{array} \right)
 \CC \left(n,n_{12},n_{21}\right)
\end{array}
\end{equation}
where the coefficient $\CC (n,n_{12},n_{21} )$ are given
by~\cite{Beisert:2003jj}
\be \CC (n,n_{12},n_{21})=(-1)^{(1+n_{12}n_{21})}\frac{\Gamma
\left(\frac{1}{2}(n_{12}+n_{21})\right) \Gamma
\left(1+\frac{1}{2}(n-n_{12}-n_{21})\right)}{\Gamma
\left(1+\frac{n}{2}\right)} \label{coeffbe} \ee
with $ \CC (n,0,0)=h(n/2)$. Moreover,
\begin{equation}
\begin{array}{l}
n=\sum_{i=1}^2(a^1_{(i)}+a^2_{(i)}+b^1_{(i)}
+b^2_{(i)}+c^1_{(i)}+c^2_{(i)}+c^3_{(i)}+c^4_{(i)}) \\[2mm]
n_{12}=\sum_{\alpha=1}^2 a^\alpha_{(1)}+\sum_{\dot{\alpha}=1}^2
b^{\dot{\alpha}}_{(1)}+\sum_{a=1}^4c^a_{(1)}-k-k'
-p-p'-\sum_{a=1}^4l_a\\[2mm]
n_{21}=\sum_{\alpha=1}^2 a^\alpha_{(2)}+\sum_{\dot{\alpha}=1}^2
b^{\dot{\alpha}}_{(2)}+\sum_{a=1}^4c^a_{(2)}-k-k'
-p-p'-\sum_{a=1}^4l_a
\end{array}
\end{equation}

\end{appendix}

\providecommand{\href}[2]{#2}\begingroup\raggedright\endgroup


\end{document}